\documentclass[final,3p,times]{elsarticle}
\usepackage{amsthm}
\usepackage{xcolor}

\usepackage{graphicx}
\usepackage{caption}
\usepackage{subcaption}
\usepackage{natbib}
\usepackage{enumitem}
\usepackage{appendix}
\usepackage[section]{placeins}
\usepackage{amsmath}
\usepackage{bbm}

\setcitestyle{authoryear,open={(},close={)}}

\usepackage{lscape}
\usepackage{algorithm,algorithmic}
\usepackage{lineno}

\newcommand{\floor}[1]{\left\lfloor #1 \right\rfloor}

\journal{Transportation Research Part C}

\begin{document}

\title{Significance of Low-level Controller for String Stability under Adaptive Cruise Control}


\author[1]{Hao Zhou}
\author[1]{Anye Zhou}
\author[2]{Tienan Li}
\author[2]{Danjue Chen}
\author[1]{Srinivas Peeta}
\author[1]{Jorge Laval\corref{cor1}}

\ead{jorge.laval@ce.gatech.edu}
\cortext[cor1]{Corresponding author}
\fntext[label1]{790 Atlantic Dr NW, Atlanta, GA, 30313}
\address[1]{School of Civil and Environmental Engineering, Georgia Institute of Technology, Atlanta, United States}
\address[2]{School of Civil and Environmental Engineering, University of Massachusetts Lowell
, Lowell, United States}

\begin{abstract}
Current commercial adaptive cruise control (ACC) systems consist of an upper-level planner controller that decides the optimal trajectory that should be followed, and a low-level controller in charge of sending the gas/brake signals to the mechanical system to actually move the vehicle. 
We find that the low-level controller has a  significant impact on the string stability (SS)  even if the planner is string stable: (i) a slow controller deteriorates the SS, (ii) slow controllers are common as they arise from insufficient control gains, from a "weak" gas/brake system or both, and (iii) the integral term in a slow controller causes   undesired overshooting which affects the SS. Accordingly, we suggest tuning up the proportional/feedforward gain and ensuring the gas/brake is not "weak". The study results are validated both numerically and empirically with data from commercial cars. 

\end{abstract}
\maketitle

\section{Introduction}

With the development of vehicle automation, adaptive cruise control (ACC) systems are now widely available on commercial vehicles around the world. From the perspective of traffic flow efficiency, ACC systems are expected to achieve string stability (SS) to ensure that small perturbations do not amplify upstream within a platoon of vehicles \citep{4282789, Feng2019StringSF}, i.e., speed fluctuations should be dampened rather than amplified by the followers \citep{naus2010string, zhou2005range, Zhou2020SmoothSwitchingCC}. In recent years, the SS of factory ACC systems has drawn increasing attention from the traffic flow community. Unfortunately, the factory ACC products can only be treated as "black boxes" due to the proprietary technology.  As a result, related studies in  the literature are mostly data-driven; i.e., researchers often collect the driving behavior data of factory ACC vehicles and then directly examine the SS features (e.g., \cite{makridis2021openacc,li2021ACC} ) or indirectly study the features by calibrating a model (e.g., \cite{gunter2019model,gunter2020commercially,shi2021empirical}) .   
For example, \cite{gunter2020commercially} found that most market ACC systems are string unstable.  \cite{li2021ACC} show that factory ACCs can amplify or dampen an oscillation (and overshoot or undershoot after the oscillation), depending on the ACC headway setting, speed level, and leader trajectory. \cite{shi2021empirical} illustrate similar findings in that a large headway setting produces better SS while a small one usually induces perturbations to grow. \cite{li2020trade} indicates that this property of factory ACC suggests a trade-off between SS, mobility and safety. These empirical findings of  "black box" factory ACC systems shed light on some of their key features, but lack insights from the perspectives of ACC controller design and execution.  

Although the data-driven studies have shed new light on understanding the factory ACC systems, there are some limitations.  Notably, in recent car models, ACC functionality is often provided by radar manufacturers, e.g. Bosch, Continental, who sell the units to automakers. Those ACC units usually integrate a built-in ACC algorithm with the radar module.  More specifically, an upper-planner receives the updated information from the radar and plans for the optimal trajectory, and then a low-level controller executes the trajectory by sending low-level commands (gas/brake or acceleration) to the car control interface.  
Given this design feature of ACC,  the data-driven approach is not ideal in uncovering ACC behaviors.  First, they use similar GPS devices with a sampling rate of only 0.1s, which  is larger than the typical updating interval for both radar (0.05s) and the low-level control system (0.01s) in modern cars. This means that some vehicle maneuver information is potentially lost.  Second, the data-drive approach only captures the holistic outcomes of ACC behaviors and cannot decouple the role of the upper-level planner and lower-level controller as well as their interactions.  Notably, the prevailing approach of model calibration per 
\cite{gunter2019model,li2020trade,shi2021empirical} essentially captures only the upper-level planner and assumes that the low-level controllers achieve perfect performance as desired, which is questionable in real-world driving scenarios.  In fact, \cite{li2020trade,shi2021empirical} calibrated the factory ACC using a parsimonious linear car following (CF) model 
and found that the model falls short in explaining some empirical ACC features.  This highlights the needs of going deeper into the mechanic level of ACC controller design and execution.

Fortunately, a recently open-sourced factory ACC system, Openpilot \citep{Openpilot}, here provides a feasible solution to overcome the proprietary issue. Openpilot is developed by an after-market self-driving company, Comma.ai, who aims to be to Tesla what Android is to Apple, but with self-driving technology. Besides the open-source ACC software, Comma.ai also develops an after-factory ACC development kit, Comma Two, which can connect the stock ACC unit and override its control on many regular commercial cars. Note that Openpilot is considered to be factory level because it only overwrites the built-in algorithms in the ACC unit, as it still uses the stock sensor and communicates with existing car control interface. The benefits of such open-source ACC software and after-market hardware are unprecedented. Openpilot enables researchers to obtain full access to all the controller parameters, variables and algorithms of the ACC system. It opens a new gate for analysis, design and test of our own factory ACC algorithms, which is aligned with the objective of this study.

Prior to the open-source factory ACC algorithms, the SS of ACC systems has already been extensively studied in the control area \citep{yanakiev,liang1999optimal,Liang2000StringSA}. 
The theoretical SS condition for linear CF models have been examined for years since \citep{wilson2011car}, and one can notice that it is not difficult to meet the SS condition after tuning the model parameters. For example, \cite{gunter2019model} obtained a wide region of string-stable parameters for the optimal speed velocity speed model and verified the SS through simulation.  \citep{wilson2011car} has derived the string stability region for parameters of some CF models.  
Yet, two interesting questions arise: can we directly apply those tuned models to a car and achieve string-stable performance? Why are string-stable commercial ACCs so rare in real life?

Unfortunately, the current short answer to the first question is "no" and the the second question can be explained by the lack of considerations for low-level controllers. Specifically, this study points out that the gap between SS theory and practice lies in the impact of low-level controller, which has been consistently neglected so far. 
For the field experiments in literature, researchers used either original equipment manufacturer (OEM) controller or in-house designed controllers. The OEM controller \citep{Eilbert2020PerformanceCO} directly tracks the acceleration/speed command, or torque/brake pressure generated from the upper-level planner but the algorithms are also 'black-boxes'. The in-house designed controllers use specific feedback control law to track the planning trajectories, e.g. \cite{naus2010string,6082981} used a 'MOVE' gateway to interact with the vehicle motor to achieve acceleration setpoint ; \cite{Lu2018TruckCS} applied feedback linearization control to calculate the desired torque/brake pressure command of truck actuators for desired accelerations; and \cite{Shladover2009EffectsOC} utilized the loop-shaping techniques to devise a 'speed servo' to track the planned speed. Although the in-house designed controllers demonstrate desired performance in the experiments, they are only applied to specific filed-test scenarios, it's unclear if they can be generalized to real-world driving. Additionally, most in-house designed controllers are not open-sourced either.

Correspondingly, with an emphasis on analyzing the impact of the factory low-level controller, this study first presents the open-source factory ACC algorithms, based on which we further investigate the impact of low-level controllers on SS. The findings are validated through both numerical simulations and real-car tests on a regular commercial car model, a 2019 Honda Civic.      

The contribution of this paper is two-fold: i) for the first time we have introduced an open-source factory ACC system to the literature, which allows us to develop a new experimental method to customize/test new ACC algorithms on many car models available on the market. Such a method provides promising opportunities to understand the market "black boxes". ii) we have shown that the low-level controller plays a decisive role in the SS of ACC, which has been neglected in the literature. This finding suggests a new direction to tackle the SS issue in controller design.

The remainder of the paper is organized as follows: section II introduces the factory ACC algorithms for both the planner and the low-level controller; section III investigates the impact mechanisms of low-level controllers on SS; section IV presents the numerical and empirical experiments and showcase the results; section V provides guidances on tuning low-level controllers for the better SS; section VI concludes the paper and discusses future directions.

\section{Factory ACC algorithms}\label{background}
\subsection{Pipeline for the longitudinal control in factory ACC}

\begin{figure}[!htbp]
\centerline{\includegraphics[width=0.8\textwidth]{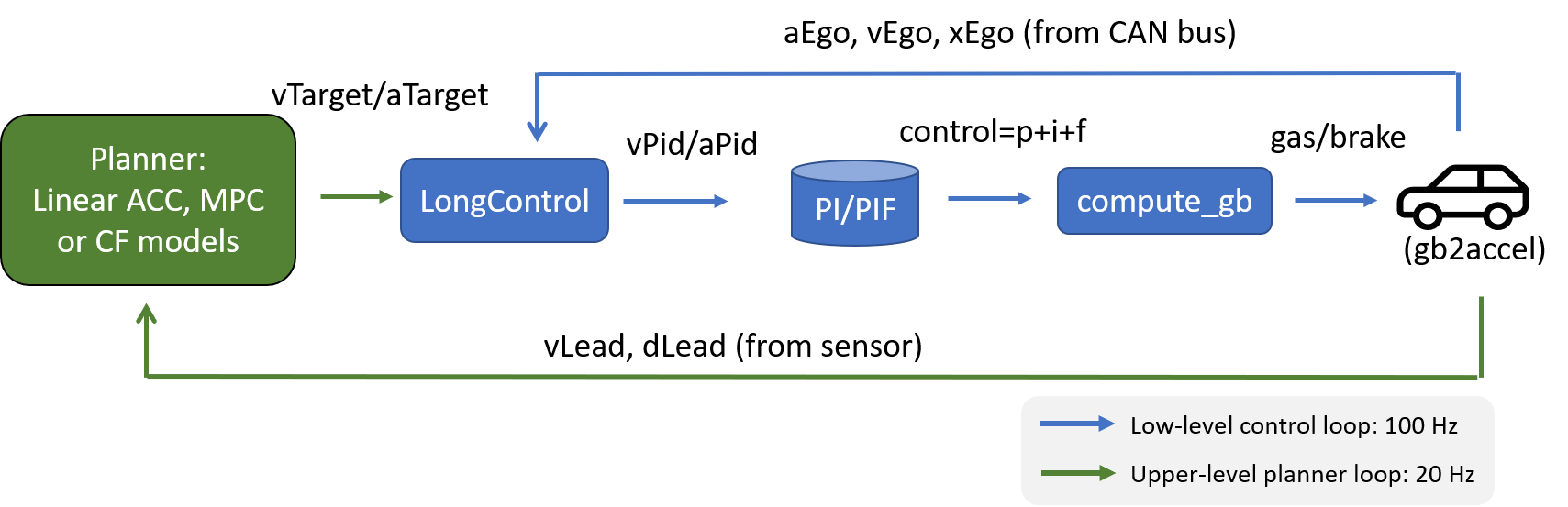}}
\caption{Pipeline for the longitudinal control in factory ACCs.}
\label{pipeline}
\end{figure}

Before diving into the detailed algorithms for the upper-planners and the low-level controllers, it is important to first understand the pipeline of the longitudinal control in a factory ACC system, as shown in Fig. \ref{pipeline}.

As mentioned earlier, a typical on-board radar runs at 20hz. It provides the lead vehicle information (e.g. the lead vehicle speed $v_{lead}$ and the headway $d_{lead}$) for the planner, with or without the sensor fusion from cameras. The planner listens to the sensors and responds to the speed and spacing changes by adjusting the target speed, $v_{\text{target}}$, or the target acceleration, $a_{\text{target}}$ for the ego vehicle. The planner model is usually a linear controller or an MPC more recently. 

The low-level controller operates to achieve the planner targets by producing the low-level commands (gas/brake), which are usually at the rate of 100hz for recent car models. Due to the higher frequency, the low-level controller forms an inner loop of 5 steps in each planning period (0.05s). As Fig.\ref{pipeline} shows, the low-level control loop consists of four major steps: i) first, the low-level control algorithm, 'LongControl', calculates the low-level setpoint, $v_{\text{pid}}$ or $a_{\text{pid}}$ or both, at each of the five control steps using the most-recent upper-level planner target $v_{\text{target}}$ or $a_{\text{target}}$ ; ii) then a proportional-integral (PI) or proportional-integral-feedforward (PIF) controller tracks the low-level setpoints and outputs the $control$ demand; iii) the $control$ demand is then fed to a $'compute\_gb'$ function which maps it to the final actuator command, i.e. a gas or brake percent $gb$ for the car; and iv) finally the gas/brake percentage is applied and moves the vehicle. The true acceleration $a_{\text{ego}}$, speed $v_{\text{ego}}$, position $x_{\text{ego}}$ of the vehicle are send back to the loop for the next step.

Here a PI or PIF controller outputs the the $control$ demand, which is further processed by a $compute\_gb$ function to generate the final $gb$ commands. Note that the $control$ can also be understood as the desired acceleration, thus the $compute\_gb$ function is essentially a mapping from the desired acceleration to the necessary gas/brake. Reversely, we refer this mapping from the applied gas/brake $gb$ to the true acceleration $a_{ego}$ as the $gb2accel$ function, which is essentially a model of the engine and brake system. 

\subsection{Planner algorithm}
Now we start introducing the planner algorithms. Chronologically, a linear ACC controller was used in the earlier versions of Openpilot, and nowadays an MPC planner replaces its role. 

\subsubsection{The factory linear planner}
We will start with the factory linear planner with a constant time headway policy (CTH). Given a desired time headway $\tau$, and the jam spacing $\delta$, the desired spacing $s_{\text{des}}$ from the front bumper of the ego vehicle to the rear bumper of the lead vehicle is calculated as:
\begin{equation}
s_{\text{des}} = \delta + \tau \cdot v_{\text{lead}}
\label{desire_d}
\end{equation}

The planner outputs a target speed $v_{\text{target}}$ to reduce the error between the true spacing $s_{\text{ego}}$ and the desired spacing $s_{\text{des}}$ by a rate $k$. 
\begin{equation}
v_{\text{target}} = (s_{\text{ego}} - s_{\text{des}}) \cdot k + v_{\text{lead}}
\label{ACC-vt}
\end{equation}
The actual $v_{\text{target}}$ will be further constrained by acceleration limits from multiple sources such as the vehicle dynamics, maximum runaway speed from the leader, or the impact of vertical and horizontal curves. 

Note that the parameter $k$ physically means how fast the planner tries to adjusts the spacing variance. In Openpilot $k$ is speed-dependent, piecewise linear 
and decreases with speed values, which is different from the constant assumptions commonly used in existing studies \citep{gunter2019model,zhou2019robust,li2020trade,shi2021empirical}. 

The desired spacing model \eqref{desire_d} indicates a linear equilibrium speed-spacing relationship, which is in accordance with the recent empirical finding from the ACC data \citep{li2021FD}. Combined with \eqref{ACC-vt}, this simple linear planner corresponds to the congested branch of Newell's simplified car-following model \citep{newell2002simplified}, which has been shown to reproduce empirical human driving data \citep{ahn2004verification,laval2014parsimonious,xu2020statistical}. In this model the ego vehicle trajectories are identical to their leader's except for a spatial and temporal translation of  $\delta$ and $\tau$, respectively.

It is also worth noting that here the factory linear ACC model uses the lead vehicle speed $v_{\text{lead}}$ to calculate the desired spacing and plan for the target speed, which is different from the tradition CTH in the literature that adjusts the desired spacing based on the ego speed $v_{\text{ego}}$. While the true motivations are unclear yet, we notice that such design enables the planner to run independently from the low-level responses such as $v_{\text{ego}}$ or $a_{\text{ego}}$. Similar design is also found in MPC-type planner, which will be introduced shortly. We conjecture that using $v_{\text{lead}}$ in the planner can also reduce the hardware communication between the ACC unit and the vehicle CAN (Controller Area Network) bus since no low-level variables need to be retrieved from the CAN bus. The $v_{\text{lead}}$ is directly measured from the radar in the ACC unit, which helps the ACC module to be self-contained. From the perspective of SS, the difference between using $v_{\text{lead}}$ and $v_{\text{ego}}$ is not trivial. We show that the factory linear ACC can be easily string stable in theory; see \ref{factory-linear-SS-proof} for the strict proof based on transfer functions.

\subsubsection{MPC planner}

MPC planners have become ubiquitous in the literature since they allow the optimization of more general objective functions while considering more refined vehicle dynamics models \citep{corona2008adaptive,naus2008explicit,li2010model, Gong2016ConstrainedOA, 7546918}. This extra computational burden comes at the cost of requiring  a professional solver \cite[e.g.][]{Acado} to run in real time. The formulation of the optimization objective and the reference trajectory are two key components for an MPC problem, as shown next.

In Openpilot the objective function $C(\hat{t})$ for the longitudinal MPC at the planning time $\hat{t}$ is defined as a weighted sum of four sub-costs, with respect to the time to collision, spacing, acceleration and jerk values:
\begin{equation}
\begin{aligned}
    C(\hat{t}) &= \sum_{\hat{t} \leq k\leq \hat{t}+T_{\text{MPC}}} w_{ttc}C_{ttc}(k)+w_{dist}C_{dist}(k) +w_{accel}C_{accel}(k) + w_{jerk}C_{jerk}(k)
\end{aligned}
    \label{MPC-cost}
\end{equation}
where $T_{\text{MPC}}$ is the optimization horizon, the tunable weights applied here are $w_{ttc}=5$, $w_{dist}=0.1$, $w_{accel}=10$, $w_{jerk}=20$. The four sub-costs are defined below: 
\begin{align}
    &C_{ttc} = \exp\{\frac{s_{\text{des}}-s_{\text{ego}}}{(\sqrt{v_{\text{ego}} + 0.5} + 0.1)/0.3}\}-1 \nonumber \\
    &C_{dist} = \frac{s_{\text{des}}-s_{\text{ego}}}{0.05v_{\text{ego}} +0.5} \nonumber \\
    &C_{accel}  = a_{\text{ego}} (0.1v_{\text{ego}}+1) \nonumber \\
    &C_{jerk}  = j_{ego}(0.1v_{\text{ego}}+1) \nonumber
\end{align}
where $j_{ego}$ is the jerk of ego vehicle (i.e., derivative of acceleration). The desired spacing $s_{\text{des}}$ for the MPC controller is defined as:
\begin{equation}
    s_{\text{des}} = v_{\text{ego}}\cdot \tau -(v_{\text{lead}}-v_{\text{ego}})\cdot \tau + \frac{v_{\text{ego}}^2- v_{\text{lead}}^2}{2B} 
    \label{mpc-desireD}
\end{equation}
where $B$ is the maximum deceleration rate. 

The reference trajectory is the estimated lead vehicle trajectory in the prediction horizon $T_{\text{MPC}}$ (2.0s). Let $dt_p$ (0.2s) denote the discrete time step  for $T_{\text{MPC}}$, the future lead trajectory is estimated using the dynamics model in 5 (a-d), which assumes a decaying acceleration of the lead vehicle with time parameter $\tau$ (1.5s). Starting from $t_0 = \hat{t}$, the lead vehicle trajectory is updated as follows:
\begin{subequations}
\begin{align}
    a_{\text{lead}} & := a_{\text{lead}}(t_0) \exp(-\tau \hat{t}^2/2) \\
    x_{\text{lead}} &  := x_{\text{lead}} + v_{\text{lead}} \cdot dt_p \\
    v_{\text{lead}} &  := v_{\text{lead}} + a_{\text{lead}} \cdot dt_p  \\
    \hat{t} &  := \hat{t}+dt_p 
    \label{lead-dynamic}
\end{align}
\end{subequations}

By solving the MPC problem \eqref{mpc-form} at each time step $\hat{t}$, we can obtain the $a_{\text{target}}(\hat{t} + dt_p)$ for the next MPC planning step:
\begin{equation}
\begin{aligned}
    \min_{a_{\text{target}}} \quad & C(\hat{t}) \\
    \textrm{s.t.} \quad &
    \text{Dynamics for the ego vehicle similar to Eqn. 5(b)(c)(d)} \\
    & \text{Predicted dynamics for the lead car in (5) } \\
    & v_{\text{target}} \ge 0
    \label{mpc-form}
\end{aligned}
\end{equation}

It is worth noting that the MPC planner can also use the predicted ego vehicle trajectory as the reference, which is more straightforward but probably needs a machine learning model. For more details, readers are referred to \cite{zhou-mpc3} which introduces a vision-based neural network model that outputs a 2s future ego vehicle trajectory.

Recall that a linear controller outputs a desired speed value $v_{\text{target}}$, which is all a low-level PI controller needs. However, it is a different case for a low-level PIF controller since it needs both $v_{\text{pid}}$ and $a_{\text{pid}}$. Consequently, the MPC planner involves two more variables $v_{\text{start}}(\hat{t})$ and $a_{\text{start}}(\hat{t})$ in the planning process,  which are updated using the following rule:
\begin{align}
a_{\text{start}} &\leftarrow a_{\text{start}} + d\hat{t}/dt_p (a_{\text{target}}-a_{\text{start}})\\
v_{\text{start}} &\leftarrow v_{\text{start}} + d\hat{t} (a_{\text{target}} + a_{\text{start}})/2
\label{aStart and vStart}
\end{align}
where $d\hat{t}$ denotes the time step for the planner to update (0.05s). 
The new variables $v_{\text{start}}$ and $a_{\text{start}}$ will serve as surrogate variables for $a_{\text{ego}}$ and $v_{\text{ego}}$ from the low-level responses. The reason for using those surrogate variables is similar to what we discussed earlier in the linear factory planner  (recall that the factory linear factory planner adopts $v_{\text{lead}}$ rather than $v_{\text{ego}}$), which is to separate the planning process from the direct low-level measurements $a_{\text{ego}}$ and $v_{\text{ego}}$. Shortly we will show how the low-level controller incorporates these planning variables $v_{\text{start}}$ and $a_{\text{start}}$. 

\subsection{Low-level controller}
The pipeline in Fig.\ref{pipeline} shows that the first step in the low-level control loop is to process the planner targets and calculate the corresponding low-level setpoints. Note that the low-level setpoints are the true references for the car to track, therefore it is important to understand how the algorithms update them using the planning targets. 


If the planner gives a target speed $v_{\text{target}}$, the low-level controller usually uses a speed setpoint $v_{pid}$ and
a PI controller to track it. Let $\hat{t}$ denote the planning time when the sensor and planner get updated every 0.05s, and $t$ is the low-level control time with step size $dt=0.01s$. The algorithm to compute $v_{\text{pid}}$ for the PI controller is shown in Algorithm 1 with initial condition $v_{\text{pid}}(0)=v_{\text{ego}}(0)$.

\begin{algorithm}
 \caption{Low-level controller algorithm for $v_{\text{pid}}$ with a linear planner}
 \begin{algorithmic}[1]
 \renewcommand{\algorithmicrequire}{\textbf{Input:}}
 \renewcommand{\algorithmicensure}{\textbf{Output:}}
 \REQUIRE  most recent $v_{\text{target}}$; $a_{max}$, $a_{min}$; a constant overshoot allowance $oa = 2.0$.
 \ENSURE  $v_{\text{pid}}$
 \\ \textbf{Low-level iteration to update $v_{\text{pid}}$}
 \IF {$v_{\text{pid}}>v_{\text{ego}}+oa$ and $v_{\text{target}}<v_{\text{pid}}$}
  \STATE $v_{\text{pid}} \leftarrow \max(v_{\text{target}}, v_{\text{ego}}+oa)$
  \ELSIF{$v_{\text{pid}}<v_{\text{ego}}-oa$ and $v_{\text{target}}>v_{\text{pid}}$}
  \STATE $v_{\text{pid}} \leftarrow \min(v_{\text{target}}, v_{\text{ego}}-oa)$
  \ENDIF
  \IF{$v_{\text{target}}>v_{\text{pid}}+a_{max} \cdot dt$}
  \STATE $v_{\text{pid}} \leftarrow v_{\text{pid}} + a_{max} \cdot dt $
  \ELSIF{$v_{\text{target}}<v_{\text{pid}}+a_{min} \cdot dt$}
  \STATE $v_{\text{pid}} \leftarrow v_{\text{pid}} + a_{min} \cdot dt $
  \ELSE
  \STATE $v_{\text{pid}} \leftarrow v_{\text{target}}$
  \ENDIF
\STATE update $v_{\text{ego}} \leftarrow v_{\text{ego}}$ from CAN bus

\label{ACC-vPid}
 \end{algorithmic} 
 \end{algorithm}
\FloatBarrier

In short, the above algorithm 1 shows the current $v_{\text{pid}}(t)$ updates itself by moving towards the given planner target $v_{\text{target}}(\hat{t})$ at a maximum rate but bounded by constraints that include acceleration limits.

If the upper-planner outputs the upper-level desired acceleration $a_{\text{target}}$, the low-level control loop usually outputs the acceleration setpoint $a_{\text{pid}}$ along with $v_{\text{pid}}$. The acceleration setpoint is designed for the feedforward term in a PIF controller. In Algorithm 2, we show the algorithm for computing $a_{\text{pid}}$ and $v_{\text{pid}}$ using the $a_{\text{target}}$, $v_{\text{start}}$ and $a_{\text{start}}$ from a MPC longitudinal planner. Initially, we reset $v_{\text{start}}(0)=v_{\text{ego}}(0)$ and $a_{\text{start}}(0) = a_{\text{ego}}(0)$.

\begin{algorithm}[!htbp]
 \caption{Low-level controller algorithm for $a_{\text{pid}}$, $v_{\text{pid}}$ with an MPC planner}
 \begin{algorithmic}[1]
 \renewcommand{\algorithmicrequire}{\textbf{Input:}}
 \renewcommand{\algorithmicensure}{\textbf{Output:}}
 \REQUIRE most recent $a_{\text{target}}$, $a_{\text{start}}$, $v_{\text{start}}$
 \ENSURE  $a_{\text{pid}}$ and $v_{\text{pid}}$
\\ \textbf{Low-level loop for $v_{\text{pid}}(t), a_{\text{pid}}(t)$ for $t \in \{\hat{t}, \hat{t}+0.01... \hat{t}+d\hat{t} \}$:}
\STATE {$dt=t-\hat{t}$} 
\STATE $a_{\text{pid}}(t) = a_{\text{start}} + dt (a_{\text{target}}-a_{\text{start}})/ dt_p$
\STATE $v_{\text{pid}}(t) = v_{\text{start}} + dt (a_{\text{pid}}(t) + a_{\text{start}})/2$
\label{mpc-apid}
\end{algorithmic} 
\end{algorithm}
\FloatBarrier

Now we introduce how the PI/PIF controllers track those low-level setpoints.
Define the true speed at each control time $t$ as $v_{\text{ego}}$, the speed error as $e(t) = v_{\text{pid}}(t) -v_{\text{ego}}(t)$, the formulation of PI/PIF  $control$ input is expressed as:

\begin{equation}
control = k_p \cdot e(t) + k_i \cdot \int_{0}^t e(\tau) d\tau + k_f \cdot a_{\text{pid}}
\label{PI_eq}
\end{equation}
where $k_p, k_i$ and $k_f$ correspond to the control gain for the proportional(P), integral(I) and feedforward(F) terms. Note that the control gains are speed-dependent in Openpilot, where $k_p=k_p(v)$ and $k_i=k_i(v)$. Similar to the parameter $k$ in \eqref{ACC-vt}, P and I gains are also piecewise linear and smaller at higher speeds. The $k_f$ equals to 1 in theory of the feedward control. The first two terms in \eqref{PI_eq} construct a PI controller, and a feedforward term is added to form a PIF controller if the acceleration setpoint is available. 

The $compute\_gb$ calculates the desired gas/brake for the $control$ demand from \eqref{PI_eq}:
\begin{equation}
gb = compute\_gb (control, v_{\text{ego}})
\label{accel2gb}
\end{equation}
where $gb\in [-1,1]$ because the brake is negative. Recalling that $control$ is the desired acceleration, to obtain the corresponding $gb$, the $compute\_gb$ must be in accordance with the true engine/brake performances of the vehicle, aka the $gb2accel$ function:
\begin{equation}
a_{\text{ego}} = gb2accel(gb,v_{ego})
\label{gb2accel}
\end{equation}

The $gb2accel$ function for a specific car model can be fit from its driving data. One can test several $gb$ values under different speed levels and record the produced true accelerations. Then a linear function or neural networks can be used to fit the model. We will later show on many recent car models, the $gb2accel$ is approximately a simple scaling function of the $gb$ only, for example, $a_{\text{ego}} = 3 gb$, and accordingly the $compute\_gb$ function should share the same scale factor in the denominator, e.g. $gb = a_{\text{ego}}/3$. 

We consider the $compute\_gb$ is perfect if the estimated $gb$ always generates the same amount of the true acceleration $a_{ego}$ as the $control$ demands: 
\begin{equation}
    control = a_{\text{ego}}
    \label{controlequal}
\end{equation}
which holds obviously when $gb2accel$ and $compute\_gb$ are both simple scaling functions with the reciprocal parameters, e.g 3 and 1/3.

Since the scale factor in $gb2accel$ is usually unknown to us, the $compute\_gb$ can be designed to have smaller or larger scale factors which causes the true acceleration overshoot or undershoot the $control$ demand. To further study their impacts, we define the actuator system is "strong" if $a_{\text{ego}}>control$, or "weak" if $a_{\text{ego}}<control$.

\section{Impact mechanisms of the low-level controller} \label{methodology}

As illustrated by the ACC pipeline, for a given string stable planner, the execution of its desired trajectory can be largely affected by the low-level controller, either by the control module the actuator, or both. In this section we will formulate the impacts of the low-level controller on SS. Later, in the next section we will provide numerical and empirical evidence.  

\subsection{A fast/slow PI controller and its impact on SS}

The dynamics of a linear planner as \eqref{desire_d} and \eqref{ACC-vt} leads to:
\begin{equation}
\begin{aligned}
    \dot{s}_{lead} & = v_{\text{lead}} - v_{\text{ego}} = v_{\text{rel}} \\
    \dot{s}_{des} &=  \tau \dot{v}_{lead} \\ 
    \dot{v}_{lead} & = a_{\text{lead}}
    \label{ACC}
\end{aligned}
\end{equation}

Substituting \eqref{ACC} into \eqref{ACC-vt}, we have: 
\begin{equation}
    \dot{v}_{target} = k (v_{\text{lead}}-v_{\text{ego}}) + (1-k  \tau) a_{\text{lead}}
    \label{v_target_rate}
\end{equation}

To investigate the impact on SS, we evaluate the maximum changes in the ego vehicle target speed, namely $\Delta v_{\text{target}}$, which can be calculated by tracking \eqref{v_target_rate} from the time $T_1$ when the ego vehicle starts to react to the lead vehicle, to the time $T_1+\Delta T$ when the ego vehicle
starts to reduce the overshoot/undershoot; see Fig.\ref{integral_overshooting}:
\begin{align}
    \Delta v_{\text{target}} & = v_{\text{target}}(T_1+\Delta T)- v_{\text{target}}(T_1) \nonumber \\
    & = \sum_{T_1\le \hat{t} \le T_1+\Delta T} k (v_{\text{lead}}(\hat{t})-v_{\text{ego}}(\hat{t})) d\hat{t} + \sum_{T_1\le \hat{t} \le T_1+\Delta T}(1-k  \tau) a_{\text{lead}}(\hat{t}) d\hat{t} \nonumber \\
    & = (1-k  \tau) \overline{a}_{lead} \Delta T+k \sum_{T_1\le \hat{t} \le T_1+\Delta T} v_{\text{rel}}(\hat{t})  d\hat{t}  
    \label{analysis}
\end{align}
where $\overline{a}_{lead}$ is the average acceleration for the lead vehicle during $\Delta T$. Since $v_{\text{lead}}$ and $a_{\text{lead}}$ do not change with the follower, the magnitude of the speed change for the ego vehicle will only depend on $v_{\text{rel}}$. Also note that $1-k \tau>0$ for typical values of the gain $k$ and desired time headway $\tau$.
 
Then we subtract the lead vehicle speed change during $\Delta T$, to yield a ``SS index'' for the linear planner, namely  $I_{linear}$:
\begin{align}
   I_{linear} = &(|\Delta v_{\text{target}}|-|\Delta v_{\text{lead}}|)/|\Delta v_{\text{lead}}| \nonumber \\
   & = \frac{k}{|\Delta v_{\text{lead}}|} (\sum_{T_1\le \hat{t} \le T_1+\Delta T} |v_{\text{rel}}(\hat{t})| - \tau |\overline{a}_{lead}| \Delta T) \nonumber \\
   & = - k \tau + \frac{k}{|\Delta v_{\text{lead}}|} \sum_{T_1\le \hat{t} \le T_1+\Delta T} |v_{\text{rel}}(\hat{t})| d\hat{t}
    \label{SS-index-ACC}
\end{align}
which can be applied to both acceleration and deceleration cases.
It can be seen that a smaller $I_{linear}$ is desirable as it corresponds to better SS. If $I_{linear} > 0$ the speed change is amplified from leader to follower, yielding a string unstable case. Using \eqref{SS-index-ACC}, the impact of the low-level controller can be understood as follows: for a fixed planner, a slow-response controller moves the follower more slowly, increasing both the relative speed $v_{\text{rel}}$ and the stabilization time $\Delta T$. A faster controller can track the planned speed faster, reduces the speed amplification $\Delta v_{\text{target}}$ and thus improves SS. 

\subsection{A fast/slow PIF controller and its impact on SS}

Similarly, for the PIF controller working with the MPC planner, we track the speed change of $v_{\text{start}}$ to study SS since it is the surrogate variable for $v_{\text{ego}}$. Setting  $a_{\text{start}}=0$ in \eqref{aStart and vStart} the $a_{\text{start}}$ sequence can be simplified using only $a_{\text{target}}$ from the MPC planner, which leads to:

\begin{equation}
a_{\text{start}}(\hat{t}) = \sum_{n = 0} ^{\floor{\hat{t}/d\hat{t}}-1} \frac{d\hat{t}}{d t_{p}} (1-d\hat{t}/d t_{p})^{\floor{\hat{t}/d\hat{t}}-1-n} a_{\text{target}}(n d\hat{t})
\label{vstart-iteration}
\end{equation}
where $\floor{\hat{t}/d\hat{t}}$ is a floor function to calculate index for planning step of $\hat{t}$ starting from zero. Substitute \eqref{vstart-iteration} into \eqref{aStart and vStart}, the changes of $v_{\text{start}}$ can be described using $a_{\text{target}}$ from the planner:

\begin{equation}
2/d\hat{t} \cdot \Delta v_{\text{start}} = \sum_{T_1\le \hat{t} \le T_1+\Delta T} a_{\text{target}}(\hat{t}) + \sum_{n = 0} ^{\floor{\hat{t}/d\hat{t}}} \frac{d\hat{t}}{dt_{p}} (1-\frac{d\hat{t}}{dt_{p}})^{\floor{\hat{t}/d\hat{t}}-1-n} a_{\text{target}}(n d\hat{t})
\label{SS-aTarget}
\end{equation}

Correspondingly, the SS index $I_{mpc}$ given a MPC planner can be derived as:
\begin{align}
   I_{mpc}&= \frac{|\Delta v_{\text{start}}|-|\Delta v_{\text{lead}}|}{|\Delta v_{\text{lead}}|}\nonumber \\
   &= -1 + \sum_{T_1\le \hat{t} \le T_1+\Delta T} |a_{\text{target}}(\hat{t})| \frac{d\hat{t}}{2|\Delta v_{\text{lead}}|} 
    + \sum_{n = 0} ^{\floor{\hat{t}/d\hat{t}}} \frac{d\hat{t} (1-d\hat{t}/dt_p)^{\floor{\hat{t}/d\hat{t}}-1-n}}{dt_p|\Delta v_{\text{lead}}|}  |a_{\text{target}}(n d\hat{t})| 
    \label{mpc-index}
\end{align}
Similarly to the linear planner, a smaller $I_{mpc}$ means better SS, and slower low-level controller underminses SS.
Specifically, from the MPC cost function \eqref{MPC-cost} and \eqref{mpc-desireD}, we concluded that a slow low-level controller can make $s_{\text{ego}}$ deviate more from $s_{\text{des}}$, which consequently induces larger cost and triggers the MPC planner to produce larger $|a_{\text{target}}(n d\hat{t})|$ to rectify $s_{\text{ego}}$ and $v_{\text{ego}}$ to desired values, thus increases the $I_{mpc}$ to the detriment of SS.

\subsection{Integral term and its impact on SS}
\begin{figure}[htbp]
    \centering
    \includegraphics[width=0.75\textwidth]{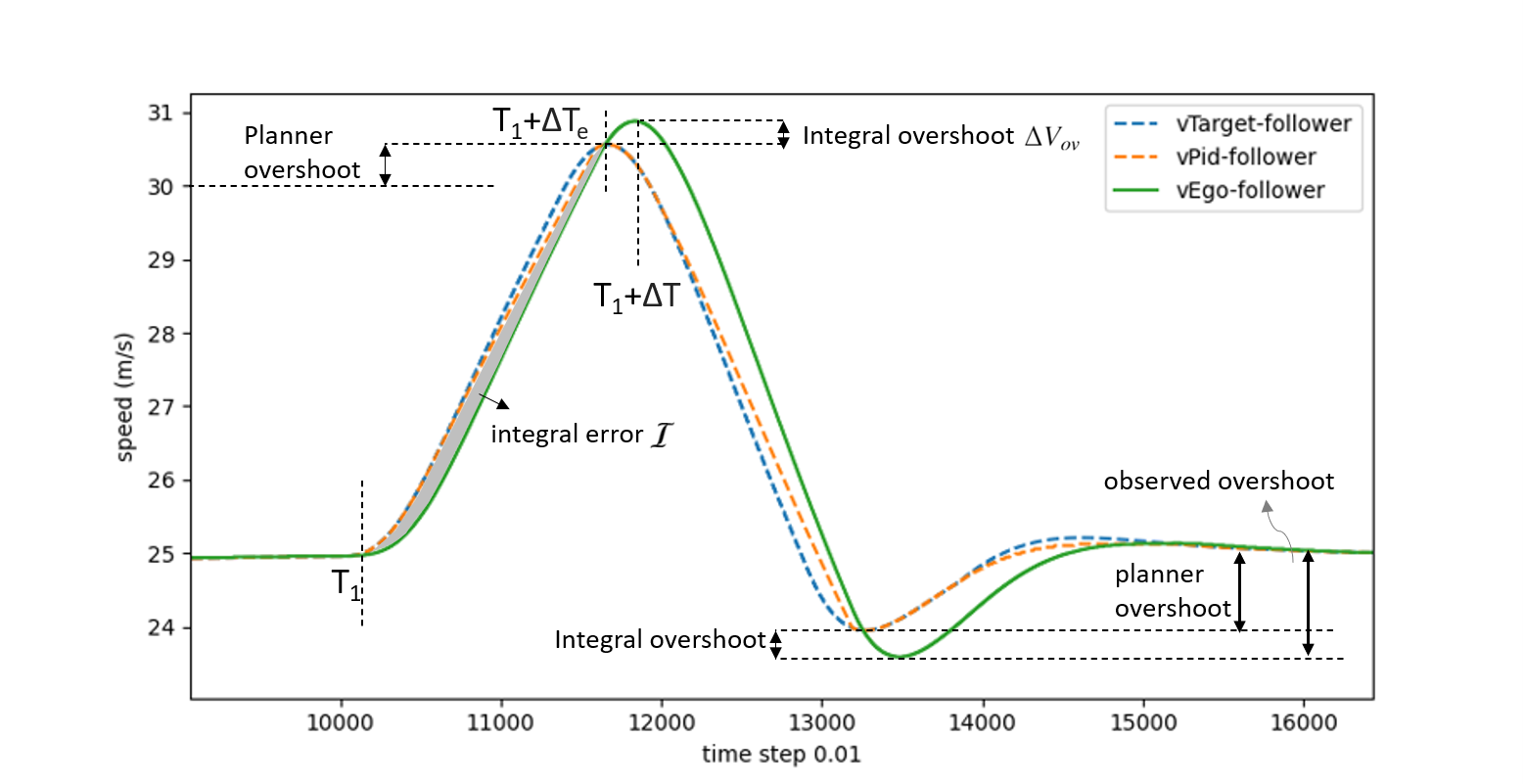}
    \caption{Integral accumulation in a PI controller during an oscillation: the lead vehicle changes its speed from 25 m/s to 30 m/s and then reverts to 25 m/s. Starting at time $T_1$, the follower speed $v_{\text{ego}}$ moves towards the setpoint $v_{\text{pid}}$ and reaches it for the first time at $T_1+\Delta T_e$, where $v_{\text{ego}}(T_1+\Delta T_e) = v_{\text{pid}}(T_1+\Delta T_e)$ and $v_{\text{ego}}(t) < v_{\text{pid}}(t)$ for $ T_1<t< T_1+ \Delta T_e < T_1+\Delta T$ in the acceleration case.}
    \label{integral_overshooting}
\end{figure}

The role of the integral term in \eqref{PI_eq} is to accelerate the object towards the setpoint, and it has been widely adopted to eliminate the residual steady-state error that occurs with a pure P or F controller. For example, a car needs integral control to generate extra gas to climb a hill. However, since the integral term responds to accumulated errors from the past, it can cause the present value to overshoot the setpoint value \citep{willis1999proportional}. For example, the shaded area in Fig.\ref{integral_overshooting} is followed by a speed overshoot. Note that the integral overshoot happens after the $v_{\text{ego}}$ catches up with $v_{\text{pid}}$. In the real world the observed overshooting \citep{li2021ACC} may come from two different sources, planner or low-level controller; see Fig.\ref{integral_overshooting} for the speed overshooting at the end of an oscillation. We argue the overshooting of $v_{\text{pid}}$ compared to the lead speed is mainly due to the planner in response to the extra spacing caused by the slow low-level controller. The additional overshooting of $v_{\text{ego}}$ compared to $v_{\text{pid}}$ mainly results from the integral accumulation. 

Now we investigate the impact of the integral error, namely $\mathcal{I}$, on SS. To this end, for tractability we will derive an upper bound for the accumulated integral error by assuming a P-only controller: 
\begin{equation}
    control \approx k_p (v_{\text{pid}}-v_{\text{ego}})
    \label{sim-dynamics}
\end{equation}

Recall that a "perfect" pair of $compute\_gb$ and $gb2accel$ functions means $a_{ego} = control$. Then, \eqref{sim-dynamics} can be written in discrete-time as follows:
\begin{align}
    v_{\text{ego}}(t+1) = v_{\text{ego}}(t) +  k_p (v_{\text{pid}}(t+1)-v_{\text{ego}}(t)) dt \quad \text{for} \quad t=1,2,3,...,
    \label{dynamics-pid}
\end{align}

Combining \eqref{dynamics-pid} with the initial condition $v_{\text{pid}}(0) = v_{\text{ego}}(0)$, we obtain $v_{\text{ego}}(n)$ at the time step $n$:
\begin{equation}
    v_{\text{ego}}(n) = (1+k_p dt) (1-k_p dt)^{n} v_{\text{ego}}(0) + k_p dt \Sigma_{i=1} ^n  (1-k_p dt)^{n-i} v_{\text{pid}}(i)\quad \text{for} \quad i=1,2,...,n-1
    \label{P-impact}
\end{equation}

Now we can quantify the tracking error $e(n)=v_{\text{pid}}(n)-v_{\text{ego}}(n)$ at time step $n$:
\begin{align}
    e(n) =  (1-k_p dt)v_{\text{pid}}(n)-(1+k_p dt)(1-k_p dt)^{n} v_{\text{ego}}(0) - k_p dt\Sigma_{i=1} ^{n-1} (1-k_p dt)^{n-i} v_{\text{pid}}(i) 
    \label{p-error}
\end{align}

Typical values for $k_p$ and  $dt$ are 1 and 0.01s, respectively. Thus $(1-k_p dt)^{n}$ tends towards zero for large $n$, making the second term $(1+k_p dt)(1-k_p dt)^{n} v_{\text{ego}}(0) \to 0$. In addition, as $k_p dt \ll 1$, the third term  $(k_p dt) \Sigma_{i=1} ^{n-1} (1-k_p dt)^{n-i} v_{\text{pid}}(i)$ is negligible comparable to $(1-k_p dt)v_{\text{pid}}(n)$. Therefore, the first term is the dominating one, which suggests the following approximation for $\mathcal{I}$:
\begin{align}
    \mathcal{I}& = \Sigma_{n=T}^{T+\Delta T_e} e(n) \approx(1-k_p dt) \Sigma_{n=T}^{T+\Delta T_e} v_{\text{pid}}(n)
    \label{over-value}
\end{align}

It can be seen the integral error $\mathcal{I}$ decreases with $k_p$ within its sensible range. This indicates that a slow controller may cause large integral error and undermines the SS.

\quad

In this section we sought to derive two indexes to measure the SS for the two major types of ACC systems, i.e. linear+PI and the MPC+PIF. Note that the SS indexes introduced here are essentially the same as the well-known Laplace-domain transfer function used in control theory, but are derived in the time domain rather than the frequency domain. A rigorous proof is simple and omitted due to space constraints. Note that the time-domain SS indexes are not designed to obtain the strict SS condition, instead they provide better physical intuitions (i.e., speed changes in car-following process) under the impact of low-level controllers, and the perturbation factors including the amplitude of lead speed change $\Delta v_{lead}$ and perturbation duration $\Delta T$.

The mathematical derivations in \eqref{SS-index-ACC} and \eqref{mpc-index} suggest the same finding; that a slow low-level controller undermines SS due to the larger relative speed and extra/insufficient spacing in the acceleration/deceleration case.  
The SS index in \eqref{SS-index-ACC} also sheds some light to the impact of the planner. Note that $k$, i.e., how fast the planner react to spacing, and the desired headway $\tau$ both affect SS as well. If the planner is more sensitive to spacing change (i.e., with a larger $k$), the ACC system is more likely to be string unstable (i.e., larger $\Delta v_{\text{target}}$ will make the planned trajectory harder to track). For the impact of $\tau$, \eqref{SS-index-ACC} indicates larger headway can benefit the SS, which is consistent with the recent empirical finding \citep{li2021ACC,shi2021empirical} from a few commercial ACC systems. 

Through analysis of the integral term, \eqref{over-value} suggests a slow low-level controller leads to larger tracking error, which further deteriorates the SS by causing greater integral overshooting. Although such effect is different from the P or F gain, they suggest the same finding: a faster controller can help improve the SS and vice versa.  

Eqn.\eqref{SS-index-ACC} and \eqref{mpc-index} indicate that SS is the outcome of the interactions between the planner and the low-level controller. Unfortunately, due to the complex coupling effects between upper-level planner and low-level controller, the closed-form solution describing the specific impact of P, I, and F gains, or the actuator performances cannot be derived. Numerical and empirical experiments will be conducted in following sections to study them in more detail.

\section{Experiments and results}
So far we have found that a slow low-level controller can undermine the SS, which can be traced back to the control gains and/or actuators. 
This section we will further study the specific impact of different control gains and the actuator performance first by numerical simulations, and then validate the findings experimentally on real cars.

\subsection{Numerical and experimental methods}
The experimental method is to run a custom Openpilot on a regular commercial vehicle, which builds upon the stock sensors and ACC interfaces of the car. The numerical method uses the same code base, but emulates the actuator (gas/brake) with a model and uses the simulation distances in lieu of the radar measurements.

For the experimental method, the hardware preparation is rather simple. We only need to connect the Comma.ai's after-market device Comma Two, to the interface of the ACC unit of the car, which will be overwriting the stock ACC algorithms and running our custom Openpilot instead. 
The driving logs from Comma Two include all the ACC-related variables, e.g. the $v_{target}$ and $k_p e(t)$, as well as the CAN bus messages of the car, such as $v_{ego}$ and $a_{ego}$. The method does not require any modifications of a regular commercial car and the full access of the ACC system allows us to dive into the details of the control algorithm and analyze its impact. A more detailed installation tutorial can be found here \citep{comma2}.

To run the ACC code numerically on computers, we have to create a virtual radar and gas/brake system. For the sensor, we replace the radar estimates $v_{lead}$ and $s_{ego}$ using the free-of-error speed and spacing from simulations. For the virtual actuator model, i.e. the $gb2accel$ function, recall that  \eqref{gb2accel} can be approximated as a simple scaling function. 
To further simulate the impact of a strong or weak actuator, one can simply manipulate the scale factor in $compute\_gb$ to change the equality in \eqref{controlequal}. 
For the other car-specific parameters such as the acceleration bound, or the control gains, we can pick the default values from any car model available in the Openpilot's pool. Here in this paper we use all default values from a Honda Civic in accordance with the field experiments. Since we focus on the longitudinal control, a tangent road is assumed where the steering angle is always zero and no lateral control is needed. We also omit other impact factors such as the grade, and the external speed disturbances.
The numerical method can run efficiently without a professional car kinematics software, it is also hazard-free which allows us to test arbitrary control gains and conduct the platoon experiments with little cost.

\subsection{Numerical results}

\subsubsection{Impact of the actuator performance}
\begin{figure}[htbp]
\centering
\begin{subfigure}[t]{0.3\linewidth}
    \includegraphics[width=\linewidth]{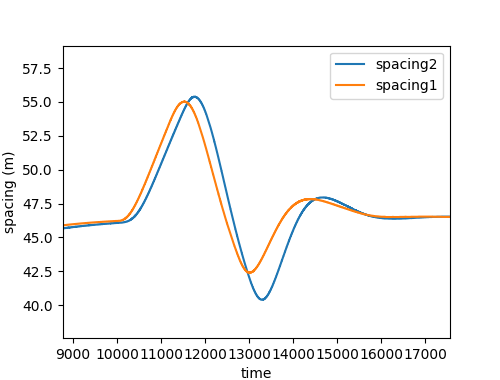}
    \caption{spacing}
\end{subfigure}%
\begin{subfigure}[t]{0.29\linewidth}
     \includegraphics[width=\linewidth]{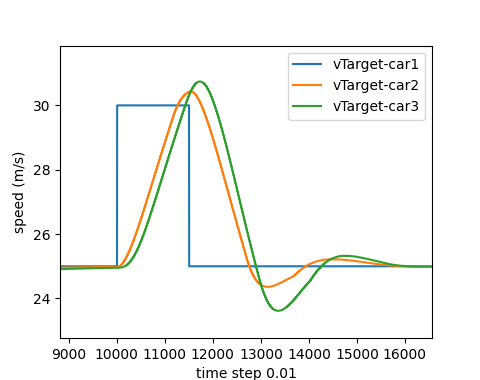}
     \caption{$v_{\text{target}}$}
\end{subfigure}%
\begin{subfigure}[t]{0.29\linewidth}
     \includegraphics[width=\linewidth]{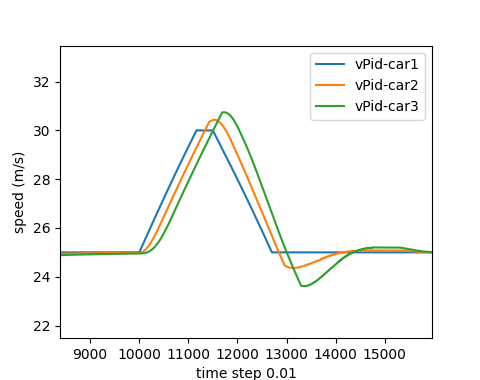}
     \caption{$v_{\text{pid}}$}
\end{subfigure}
\\
\begin{subfigure}[t]{0.3\linewidth}
     \includegraphics[width=\linewidth]{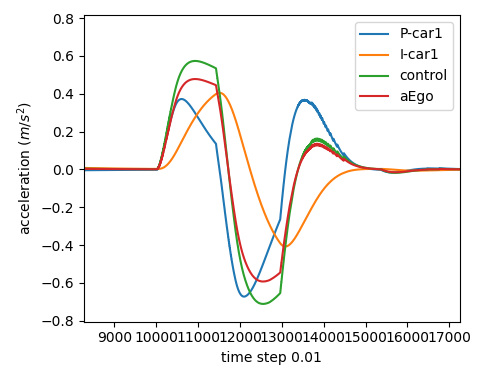}
     \caption{$control$}
\end{subfigure}
\begin{subfigure}[t]{0.31\linewidth}
     \includegraphics[width=\linewidth]{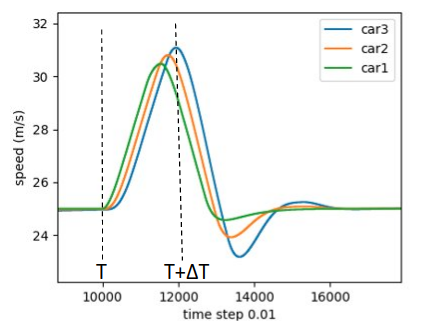}
     \caption{speed}
\end{subfigure}
    \caption{The impact of a weak actuator  on SS: the order of the figures are intended to show the cause and effect but it is a loop indeed.}
    \label{undershooting_actuator}
\end{figure}

\begin{figure}[htbp]
\centering
\begin{subfigure}[t]{0.31\linewidth}
    \includegraphics[width=\linewidth]{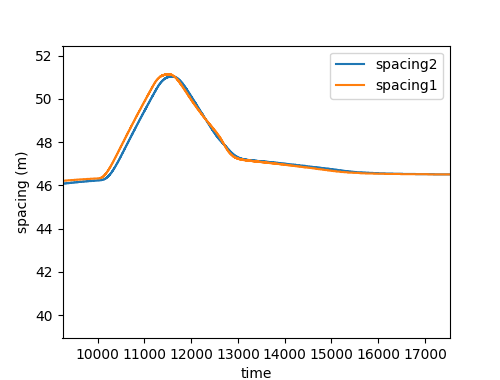}
    \caption{spacing}
\end{subfigure}%
\begin{subfigure}[t]{0.3\linewidth}
     \includegraphics[width=\linewidth]{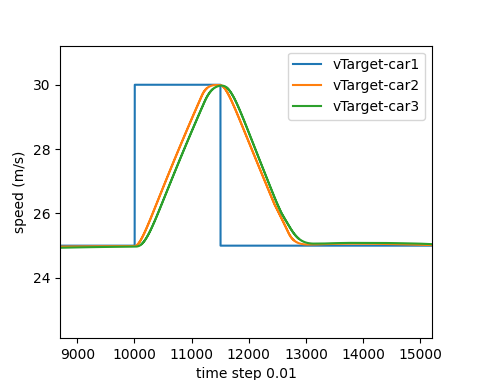}
     \caption{$v_{\text{target}}$}
\end{subfigure}%
\begin{subfigure}[t]{0.3\linewidth}
     \includegraphics[width=\linewidth]{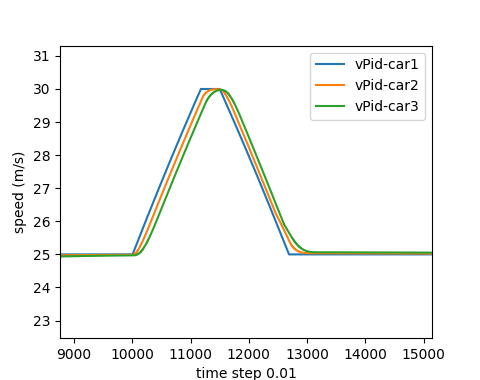}
     \caption{$v_{\text{pid}}$}
\end{subfigure}
\\
\begin{subfigure}[t]{0.285\linewidth}
     \includegraphics[width=\linewidth]{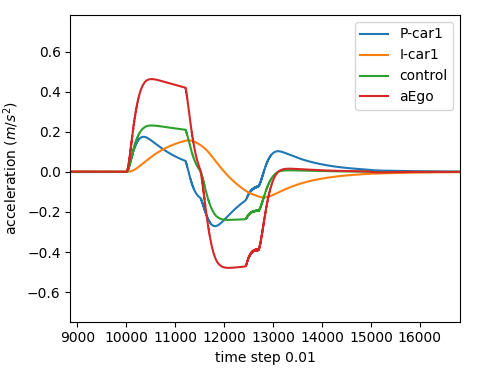}
     \caption{control}
\end{subfigure}
\begin{subfigure}[t]{0.31\linewidth}
     \includegraphics[width=\linewidth]{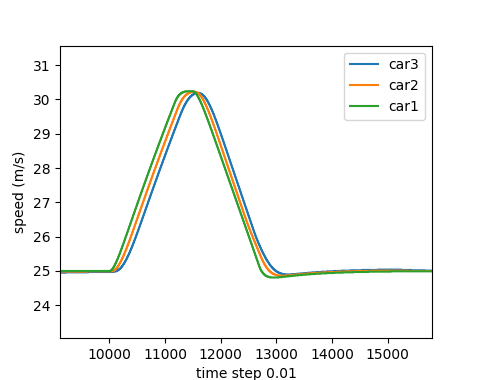}
     \caption{speed}
\end{subfigure}
    \caption{The impact of a strong actuator on SS}
    \label{overshooting_actuator}
\end{figure}

We first show numerical simulations regarding the impact of the actuator (i.e., $gb2accel$ function). Recall the true acceleration $a_{ego}$ can overshoot or undershoot the desired $control$ given a strong/weak actuator.
Now we conduct the numerical experiments of a weak actuator and a strong actuator in a linear+PI ACC system. The simulation results are shown in Fig.\ref{undershooting_actuator} and Fig.\ref{overshooting_actuator}, respectively. 

As comparison, the strong actuator (see Fig.\ref{overshooting_actuator}(e)) achieves significantly better SS than the weak actuator (see Fig.\ref{undershooting_actuator}(e)). For the speed tracking error, we see the slow controller has larger error, as indicated by the P term which is proportional to the speed tracking error (recall the first term in \eqref{PI_eq}); see blue curves in Fig.\ref{undershooting_actuator} (d) and Fig.\ref{overshooting_actuator} (d). Additionally, for the time $\Delta T$ that the follower needs to stabilize, Fig.\ref{undershooting_actuator} (e) and Fig.\ref{overshooting_actuator} suggest the slow controller needs longer time. Both findings support our interpretation of the mathematical derivations \eqref{SS-index-ACC} and \eqref{mpc-index}. 

\subsection{Impact of control gains}\label{gains}
Now we investigate the impact of different control gains. To do so, we first rule out the impact of the actuator by using the perfect actuator model such that \eqref{controlequal} holds. Then we tune the default control gains in Openpilot, which are denoted as $k_p^0$, $k_i^0$ and $k_f^0$ for P, I, F terms repsectively. Note the P and I gain are speed-dependent, i.e. $k_p^0=k_p^0(v)$ and $k_i^0=k_i^0(v)$. The feedforward gain $k_f^0=1.0$ for all speed levels.

\subsubsection{Proportional gain}
The impact of P gain is similar to that of a strong/weak actuator. For a P-only controller, Fig.\ref{integral_term} (d) shows a scaled P gain changes the $control$ in the similar way that a scaled $gb2accel$ function does. For a regular PI controller, increasing P gain also makes the low-level controller faster.Here we show the differences of the SS resulting from the small P gain and the large P gain in the PI controller, which are $k_p^0$ and $2k_p^0$, respectively. The detailed distinctions in spacing, planning variables $v_{target}$, and low-level variables $v_{\text{pid}}$ are all similar to the comparison between Fig.\ref{undershooting_actuator} and Fig.\ref{overshooting_actuator}.

\begin{figure}[htbp]
\centering
\begin{subfigure}[t]{0.4\linewidth}
    \includegraphics[width=\linewidth]{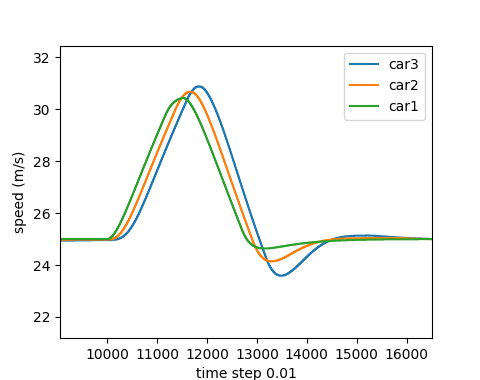}
    \caption{the small P gain: $k_p^0$}
\end{subfigure}%
\begin{subfigure}[t]{0.4\linewidth}
    \includegraphics[width=\linewidth]{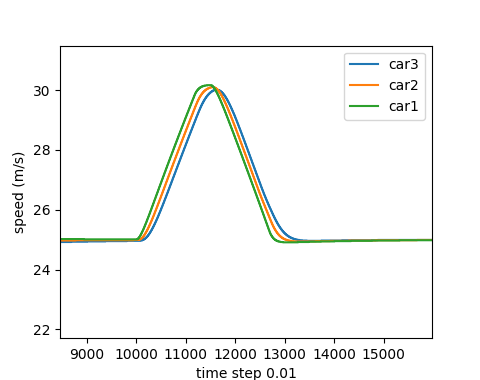}
    \caption{the large P gain: 2$k_p^0$}
\end{subfigure}%
    \caption{Impact of P gain in the PI controller on the SS}
    \label{P-gain}
\end{figure}
\FloatBarrier

\subsubsection{Integral gain}
\begin{figure}[htbp]
\begin{subfigure}[t]{0.3\linewidth}
    \includegraphics[width=\linewidth]{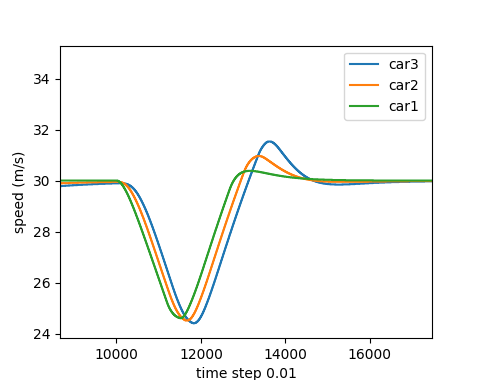}
    \caption{overshoot:P+I}
\end{subfigure}%
\begin{subfigure}[t]{0.3\linewidth}
    \includegraphics[width=\linewidth]{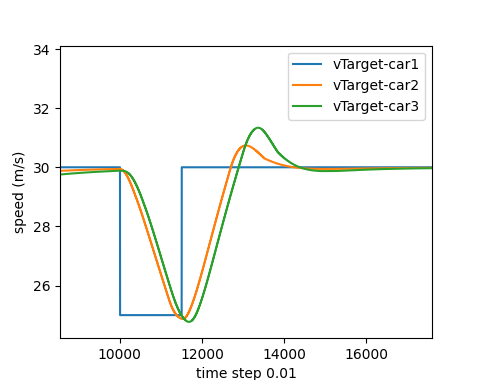}
    \caption{$v_{\text{target}}$:P+I}
\end{subfigure}%
\begin{subfigure}[t]{0.3\linewidth}
     \includegraphics[width=\linewidth]{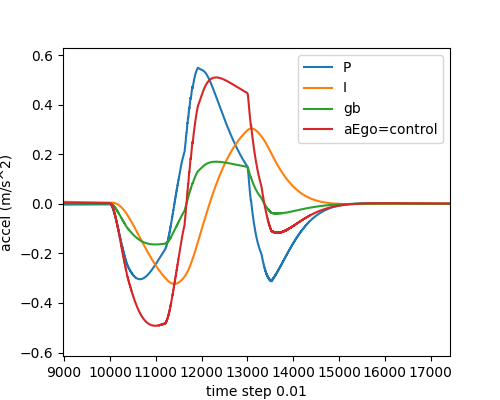}
     \caption{slow control:P+I}
\end{subfigure}%
\\
\begin{subfigure}[t]{0.3\linewidth}
    \includegraphics[width=\linewidth]{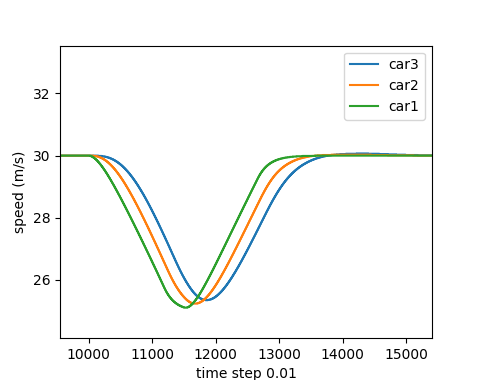}
    \caption{no overshoot:P only}
\end{subfigure}%
\begin{subfigure}[t]{0.3\linewidth}
    \includegraphics[width=\linewidth]{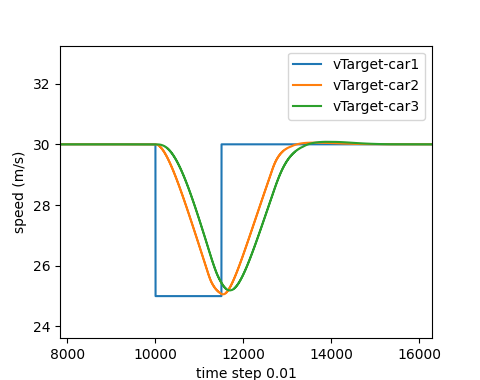}
    \caption{$v_{\text{target}}$:P only}
\end{subfigure}%
\begin{subfigure}[t]{0.3\linewidth}
     \includegraphics[width=\linewidth]{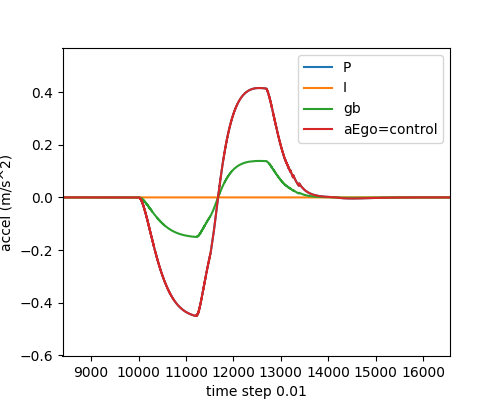}
     \caption{fast control:P only}
\end{subfigure}%
\caption{The impact of integral term on SS}
\label{integral_term}
\end{figure}

To show the impact of integral term, we compare the results from a PI controller and a pure-P controller (see Fig.\ref{integral_term}). While the P-only controller shows no overshooting and a string stable pattern, the PI controller clearly shows speed overshooting both at the peak of and the end of an oscillation. In detail, we notice the integral term is an accumulation of past error, which can sometimes have opposite sign against the current proportional term. The integral term would further cancels out part of the proportional term, in order to reduce the accumulated error (see Fig.\ref{integral_term}(c)). Thus, inappropriately-tuned integral gain could slow down the low-level controller, which deteriorates the SS.

In comparison, a P-only controller performs better in terms of SS. Unfortunately, it is infeasible to discard the integral term in practice considering its role in reducing steady errors (i.e., for desired tracking of lead vehicle). To avoid the dilemma, large P or F gain is suggested to prevent excessive tracking errors from accumulating.

\FloatBarrier

\subsubsection{Feedforward gain}
In a PIF controller, the feedforward term is $k_f \cdot a_{\text{pid}}$.
While the P and I term would wait for the error to accumulate first and then reduce it, the F term is able to generate an appropriate response in advance by incorporating the predicted acceleration, i.e. $k_f \cdot a_{\text{pid}}$, where the F gain $k_f$ is usually set to 1. The feedforward term delivers the response without the feedback errors, which usually prevails among the three control sources in the PIF controller. 

\begin{figure}[htbp]
\centering
\begin{subfigure}[t]{0.35\linewidth}
    \includegraphics[width=\linewidth]{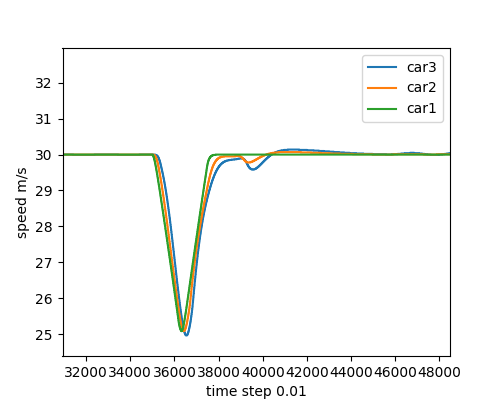}
    \caption{speed with small F}
\end{subfigure}%
\begin{subfigure}[t]{0.35\linewidth}
     \includegraphics[width=\linewidth]{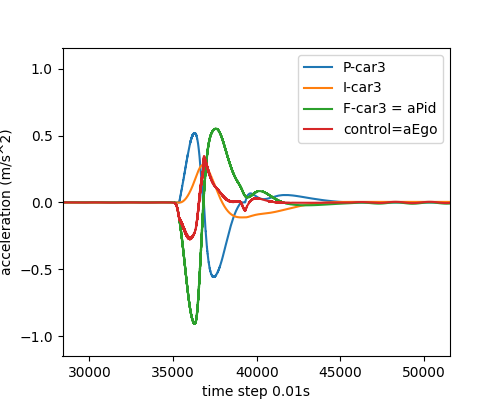}
     \caption{control with small F}
\end{subfigure}%
\\
\centering
\begin{subfigure}[t]{0.35\linewidth}
     \includegraphics[width=\linewidth]{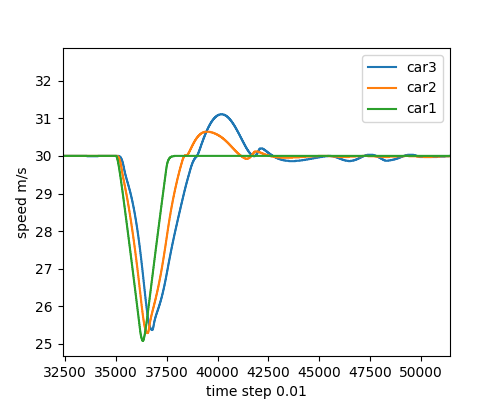}
     \caption{speed with large F}
\end{subfigure}
\begin{subfigure}[t]{0.35\linewidth}
     \includegraphics[width=\linewidth]{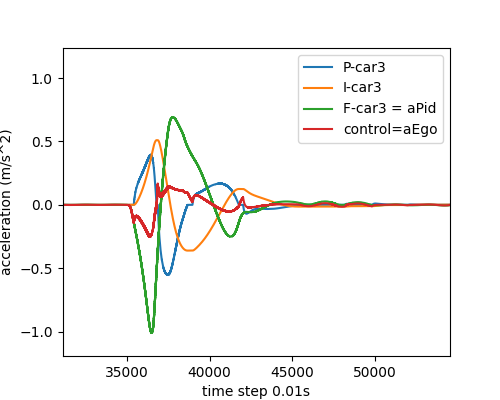}
     \caption{control with large F}
\end{subfigure}
    \caption{The impact of F gain on SS}
    \label{F-gain-impact}
\end{figure}

In simulation we run the MPC+PIF loop and conducted a comparison between a default and doubled (2.0) F gain. The results are shown in Fig.\ref{F-gain-impact}. The comparison results suggest that tuning up the F gain can also make the low-level controller faster (see Fig.\ref{F-gain-impact}(b) and (d)), which improves the SS (see Fig.\ref{F-gain-impact}(a) and (c)). We also observed that P and I terms can sometimes be contrary to the F control because $a_{\text{pid}}$ and $e = v_{\text{pid}}-v_{\text{ego}}$ can have different signs. Note that although a large F gain, such as $2 k_f^0$, brings better SS, the controller exhibits undesired oscillations at the steady state. Thus, it is suggested to gradually increase F gain with more caution to ensure the controller stability.

\subsection{Experimental results}

Now we conduct field experiments on a commercial car model to validate the aforementioned assumptions and findings. A 2019 Honda Civic is used in accordance with the default parameters chosen in simulations. The actuator model $gb2accel$ will be validated first, then we verify the impact of the integral term, and the slow/fast low-level controller on the SS.  

\subsubsection{True relationship between gas/brake and acceleration}
\begin{figure}[htbp]
\centering
\begin{subfigure}[t]{0.45\linewidth}
    \includegraphics[width=\linewidth]{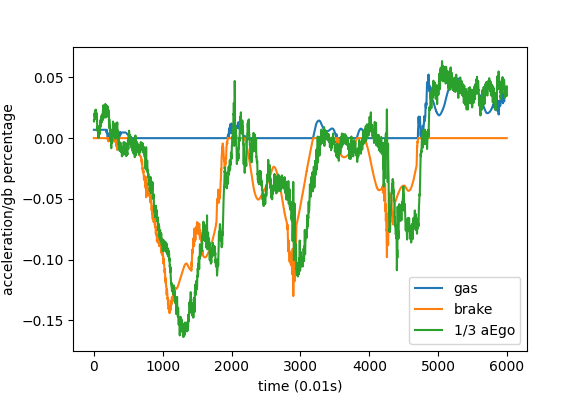}
    \caption{Toyota}
\end{subfigure}%
\begin{subfigure}[t]{0.435\linewidth}
    \includegraphics[width=\linewidth]{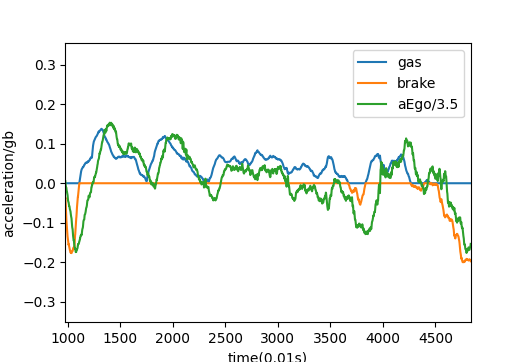}
    \caption{Honda}
\end{subfigure}%
    \caption{Relation between gas/brake and acceleration on real cars}
    \label{accel2gb_relation}
\end{figure}

Previously we assume the $compute\_gb$ and the $gb2accel$ as simple scaling functions. Now we show the empirical data collected from a 2020 Toyota Corolla and a 2019 Honda Civic in Fig.\ref{accel2gb_relation}. We can see a simple scaling function is a good fit to the relationship between gas/brake and true acceleration, though the errors tend to grow when the acceleration changes the direction, possibly due to shift in gas and brake.

The scaling functions maybe a bit surprising or even counter-intuitive because theoretically the speed should also affect the dynamics as indicated by \eqref{gb2accel}.
The simple scaling functions are devised through the special design of the car control interface on recent car models, which can accept the scaled accelerations, or $gb$ equivalently, as the input and execute the true accelerations perfectly at all speed levels. For cars equipped with this control interface, we can view the $gb$ as a scaled acceleration, rather than the true gas/brake percentages. Such scaling factor is predetermined by the car manufactures.
According to the discussions in Openpilot community \citep{compute_gb}, at least the latest car models from General Motor and Toyota are found to share this feature.


\FloatBarrier

\subsubsection{Real-car validation of the integral impact}
\begin{figure}[htbp]
\centering
\begin{subfigure}[t]{0.4\linewidth}
    \includegraphics[width=\linewidth]{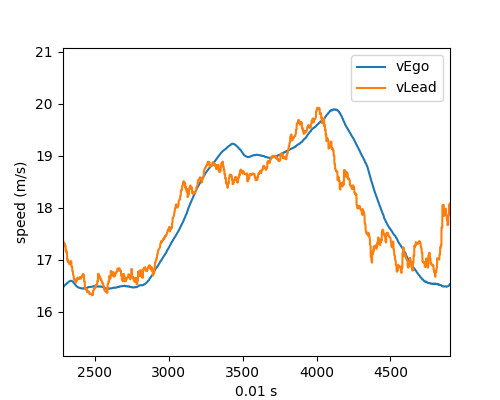}
    \caption{integral overshooting (speed) }
\end{subfigure}%
\begin{subfigure}[t]{0.4\linewidth}
    \includegraphics[width=\linewidth]{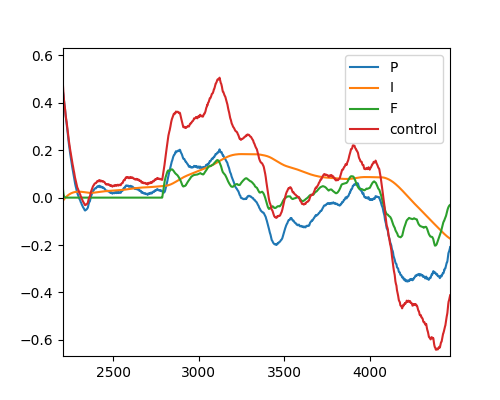}
    \caption{Integral overshooting (control)}
\end{subfigure}%
    \caption{Impact of integral overshooting on SS in a real drive}
    \label{real-windup}
\end{figure}

To validate the impact of the integral term, we conducted a real drive and compared the detailed control values from all three sources (P, I and F).  
The Fig.\ref{real-windup}(a) showcases a string unstable example where the follower overshoots the lead vehicle around step 3300, meanwhile the Fig.\ref{real-windup}(b) clearly shows that the overshoot is much due to a large and dominating integral value since both P and F terms are close to zero or even negative. Taken together, the two figures suggest the integral term accumulates when the follower accelerates to catch the leader, similar to what we have explained in Fig.\ref{integral_overshooting}. Then an immediate overshoot takes places where the opposite relationship between the I term and P/F terms is consistent with our findings from the simulations; see Fig.\ref{integral_term}(c). 

\FloatBarrier

The similar empirical observations of the overshoots have also been reported in our recent experiments on the commercial ACC systems \citep{li2021ACC}. Among the three tested car models in the experiments, two show significant overshoots while the other behave much better in preventing them. This indicates the integral overshooting could be common in commercial ACCs, but a fast low-level controller could help alleviate such effect and probably has been applied in some cars models. Since the observed overshoots can result from both the planner and the low-level controller; see Fig.\ref{integral_overshooting}; further analysis is currently conducted to enhance the empirical evidence.

\subsubsection{Real-car validation of the impact of a fast/slow low-level controller }


To verify the impact of the fast/slow tracking performance, we now test and compare two low-level controllers sharing the same planner (MPC) but with the different control gains and actuator performances. Following the experimental method in Section 4, two custom MPC+PIF branches are used to overwrite the stock ACC in a 2019 Honda Civic. Specifically, the fast low-level controller uses the control gains 1.0$k_p^0$, 0.33$k_i^0$, 1.2$k_f^0$ and the $compute\_gb$ is defined as $gb=1/3 \cdot control$ to make the actuator strong. By contrast, the slow controller adopts 0.5$f_p^0$, 0.33$k_i^0$, 1.0$k_f^0$ and the $compute\_gb$ is set to $gb=1/5 \cdot control$ such that the actuator is weak.

\begin{figure}[htbp]
\centering
\begin{subfigure}[t]{0.4\linewidth}
    \includegraphics[width=\linewidth]{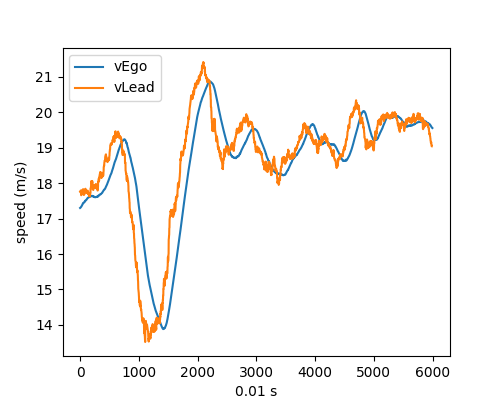}
    \caption{ a drive with the fast low-level controller }
\end{subfigure}%
\begin{subfigure}[t]{0.4\linewidth}
    \includegraphics[width=\linewidth]{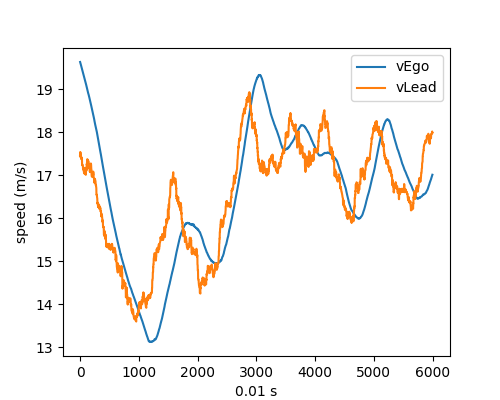}
    \caption{a drive with the slow low-level controller}
\end{subfigure}%
    \caption{Impact of a fast/slow low-level controller on SS in real drives}
    \label{real-fast-slow}
\end{figure}
\FloatBarrier

Fig.\ref{real-fast-slow} displayed the results from the two real drives using the fast/slow low-level controller. The ego vehicle was following a human-driven lead vehicle in the natural driving where the lead changes its speed occasionally on a curvy road. It is apparent that the fast low-level controller is able to dampen the lead speed changes while the slow low-level controller amplifies them. The results verified our finding that a fast low-level controller improves the SS and vice versa. 

\begin{figure}[!htbp]
\centering
\begin{subfigure}[t]{0.4\linewidth}
    \includegraphics[width=\linewidth]{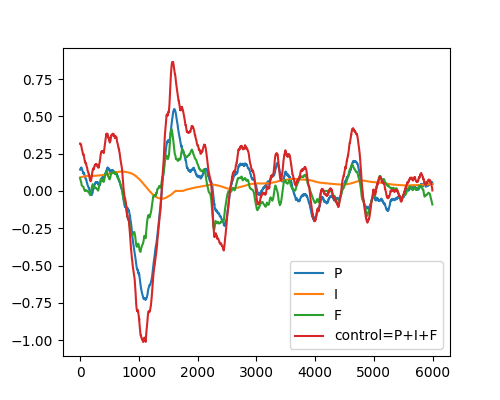}
    \caption{$P,I,F$ terms in the low-level $control$ variable}
\end{subfigure}%
\begin{subfigure}[t]{0.4\linewidth}
    \includegraphics[width=\linewidth]{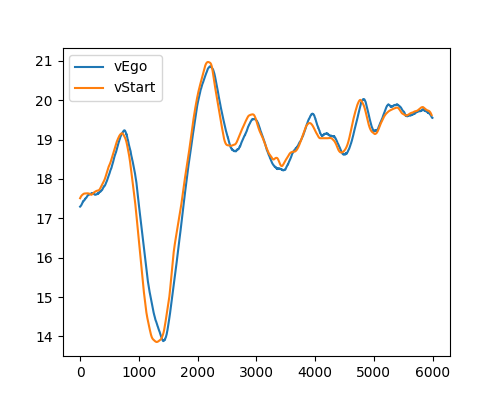}
    \caption{$v_{\text{ego}}$ follows the planning variable $v_{\text{start}}$ closely}
\end{subfigure}%
    \caption{Details of the low-level and upper-level variables in a real string-stable drive}
    \label{real-fast-detail}
\end{figure}

We further show more control details in the fast low-level controller and help explain why it improves the SS. Fig.\ref{real-fast-detail} (a) displays the detailed $P,I,F$ terms in the $control$ input during the 1-minute drive. As we pointed out earlier, the $F$  and $P$ term are expected to be consistent which makes the controller faster, rather than canceling out each other. Also a faster controller helps prevent the overshoot caused by the integral accumulation because the speed tracking error is always small and never dominates the control. Fig.\ref{real-fast-detail} (b) displays that $v_{\text{start}}$ is close to $v_{\text{ego}}$, which supports our method to use $v_{\text{start}}$ to approximate $v_{\text{ego}}$ when deriving the SS index for MPC+PIF in Section \ref{methodology}.   

\FloatBarrier

\section{Guidance on tuning low-level controllers}

The Ziegler-Nichols method \citep{Ziegler1942OptimumSF} is commonly used to obtain control gains for a PID controller. Its procedure is to first set the integral and derivative gain to zero and gradually increase the P gain until the system exhibits oscillatory behavior. Then, a look-up table provides the estimated values of control gains.  

In this study, since our control gains $k_p(v)$ are speed-dependent, we propose the following two-step method as an extension of the Ziegler-Nichols method which only considers constant control gains. The first step is to obtain  an initial feasible $k_p(v)$ that tracks the speed reference and drives the car. The second step is to fine tune $k_p(v)$ to achieve a fast low-level controller for better SS.  

To derive the initial P gains of the low-level controller, we conjecture that they should be proportional to the maximum acceleration the engine is able to output, namely  $a^*(v)$.   \cite{rakha2001vehicle} show that a straight line provides a good fit for this function. Note that a straight line fit also matches empirical data of the desired acceleration of human driven vehicles \citep{laval2014parsimonious,xu2020statistical}. 
Since $a^*(v)$ is the upper bound for the acceleration, it should satisfy:
\begin{align}
    a^*(v) &\geq \max a_{ego}(v) 
\end{align}

To understand why our conjecture should be robust, we combine \eqref{sim-dynamics} and \eqref{controlequal} to obtain: 
\begin{equation}
     k_p(v) \cdot e(t) \approx a_{ego}(v(t))
    \label{controlequalaego}
\end{equation}
Further, from Algorithm 1 we note that the maximum speed error in the low-level controller is $\max e(t)=2m/s$. Thus, the maximum acceleration generated by the low-level controller is $\max\limits_{t} a_{ego}(v) \approx 2k_p(v)$ at speed $v$, which indicates that the P gain $k_p(v)$ is a multiple of the maximum acceleration.  

The maximum acceleration in control design should be approximately equal to the true acceleration bound corresponding to the engine, which gives: \begin{align}
    k_p(v) & \le a^*(v)/2
    \label{multiple}
\end{align}

\begin{figure}[!htbp]
\centering
\begin{subfigure}[t]{0.45\linewidth}
    \includegraphics[width=\linewidth]{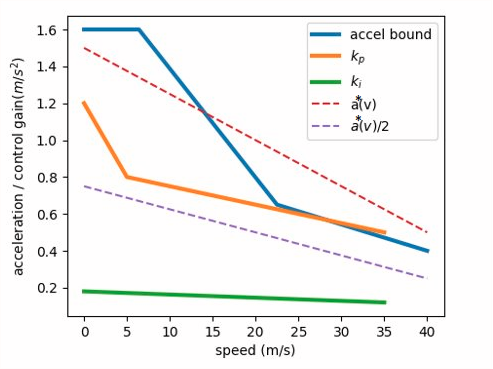}
    \caption{Default acceleration bound and control gains in Openpilot}
\end{subfigure}%
\begin{subfigure}[t]{0.45\linewidth}
    \includegraphics[width=\linewidth]{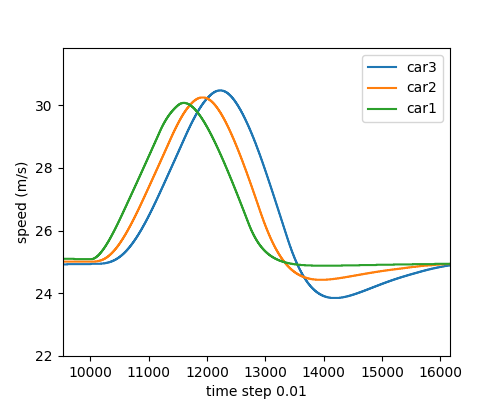}
    \caption{Performance of the initial gains from linear acceleration model}
\end{subfigure}%
    \caption{Default P and I gains in Openpilot and the initial P gain using linear acceleration model: $k_p$ and $k_i$ are the default control gains, $a^*(v)$ is the linear acceleration bound, and $a^*(v)/2$ is the proposed initial P gain }
    \label{initialP}
\end{figure}

The default acceleration upper bound used in Openpilot shown in Fig.\ref{initialP}(a) is piecewise linear and larger than the linear accelerations of human drivers. If we assume $a^*(v) = 1.0(1-v/40)+0.5$ for a regular car model, then using \eqref{multiple} we can derive the initial P gains $k_p(v) =0.5(1-v/40)+0.25$. We test the performance of such initial P gains which are feasible for driving but string unstable, using simulation (see Fig \ref{initialP}(b)). The results validate our approach to derive the initial P gains.  

For a PI controller, we still need the initial value for the I gains. A simple and common method is to determine how long it will take for an integral action to match a proportional action. For example, if the ideal time is 10 steps in the control loop, then it leads to $\Sigma_{1}^{10} k_i(v) e \approx k_p(v) e$, i.e. $k_i(v) \approx k_p(v)/10$. 

\FloatBarrier

Next, we tune the control gains to enable fast tracking performance. To this end, we suggest keeping the shape of $k_p(v)$ and only tuning a scaling factor before it, i.e. $s \cdot k_p(v)$ where $s$ should be gradually increased from 1. For tuning the controller faster, a large body of instructions, handbooks and tools \citep{handbook,tunefast} can be found online. More advanced tuning methods can also be found in the literature \citep{wang2000optimal,o2006pi,kanojiya2012optimal}. The general procedure is to gradually increase the gains while circumventing instability. The most straightforward method is trial-and-error. To verify the new control gains, the SS performance can be evaluated by running the platoon experiments using simulation as shown in Fig.\ref{P-gain}, as it is efficient and hazard-free. In the simulation experiments, the control gains are increased until the desired SS is achieved. Then, field experiments are recommended for further testing because some other aspects such as driving comfort, impact of grades, or sudden disturbances, also need to be accounted for in the real world. The SS can be evaluated from the field experiments as shown in Fig.\ref{real-fast-slow}.  

The above procedures are designed for a PI controller, based on which the initial control gains for a PIF low-level controller are easier to determine and one can apply the same method for tuning it fast. To ensure that the gas/brake is strong enough, real driving data needs to be collected for the  specific car model (see Fig.\ref{accel2gb_relation}) and fit to the actuator model in \eqref{gb2accel}, based on which the $compute\_gb$ function can be designed to ensure that the actuator is strong.

\section{Concluding comments}

This study investigates the impact of low-level controllers on string stability, which has previously been ignored in the literature. It illustrates the importance of a fast-tracking low-level controller for ensuring vehicular string stability. The study results are based on open-source factory ACC algorithms in Openpilot, Comma.ai. While stock ACC algorithms on commercial car models may vary and still remain proprietary, we conjecture that they follow a two-level control framework similar to the one introduced in this study as they share the same ACC units from the limited number of radar manufacturers. As more evidence, we note that the factory ACC algorithms presented here are able to explain some recent empirical findings from market ACC vehicles, such as the varying SS at different speed levels/ headway settings, the observed overshooting/undershooting at the end of an oscillation, and the puzzling gap between many string stable planner designs and string unstable platoons in the real world. The theory and methods proposed in this study provide promising new venues to explain more empirical ACC features or approximate the "block-box" ACC products on the market.

To encourage more experimental testing of ACC algorithms on real cars, the authors have shared a custom fork of Openpilot. This repository includes the implementation of the linear ACC+PI and MPC+PIF frameworks presented in this paper. The low-level controllers are fine-tuned for a 2019 Honda Civic and can be tested in the field by interested readers. We have added a parser to process the raw log files from Comma Two and help retrieve the ACC variables and the CAN bus signals.  
It is worth noting that the experimental method not only applies to the study of the low-level controllers; any ACC algorithm design (for example, a new planner model) can be tested in the same way.

The study formulates the control algorithms in an open-source factory ACC system. We find that the factory ACC system, whether linear ACC or MPC, does not incorporate the low-level responses (e.g. ego vehicle speed or acceleration) in the planning loop. This may be due to hardware design considerations, but it clearly distinguishes the factory planners from a regular CF model or rule-based controller in the current literature. We point out their differences here; our study can also serve as a potential catalyst for further studies using similar mechanisms to investigate the impact of these changes on safety or traffic mobility.

The study insights suggest that the factory ACC planner design is somewhat ad hoc, lacking base theory and requiring improvements, especially for the MPC planner. For example, the desired distance of the MPC in \eqref{mpc-desireD} is quite different from that in the linear ACC \eqref{desire_d}. Thereby, the motivation for the specific desired distance value is not straightforward and needs meaningful justification. Similarly, the weights for the different sub-costs in \eqref{MPC-cost} are often empirically chosen but need to be systematically linked to safety or the SS performance. These gaps between the choice of parameter values and their theoretical underpinnings provide directions for future research.  

Our comprehensive review suggests that primarily there are two major types of ACC planners, linear and MPC controllers, both of which originate from the control domain. In this context, it is surprising to note that the large body of the well-established CF models in the traffic flow domain has neither been applied or nor tested. That is, the pipeline in Fig.\ref{pipeline} suggests that any model type should work if it can output reasonable target speed or acceleration. This motivates the exploration of the promise of CF models through future research efforts and their feasibility for commercial ACC products. To enable such efforts, the authors have implemented the well-known Intelligent Driver Model (IDM) in the github repository to replace the role of a linear ACC or MPC. Future research efforts can also explore replacing the IDM model with other CF models in our shared repository and testing their performance in the field.

The self-driving industry, including ACC, has been rapidly developing in recent years. We note that ACC technologies are illustrating a trend \citep{zhou2021review} of shifting from radar-only methods to vision-assisted or even vision-only approaches. Tesla recently started to fully circumvent the use of radars and focus on the exclusive use of a pure vision Autopilot \citep{tesla-remove}. Based on the putative advantages of such an approach, we conjecture that in the future more ACC systems will adopt machine learning models (such as the deep neural networks) as their longitudinal control planners. In this context, the current theory on the SS condition only applies to linear ACC models. Hence, SS in vision-based machine learning ACC models needs more investigation, which is the focus of our ongoing work.

\section*{Acknowledgment}
This study is supported by NSF CPS grant \#1932451 and \#1826162.
The authors would like to acknowledge the warm help from Joe Pancotti and Chris Souers for their commits in Honda Bosch interface \citep{honda-bosch}. We also appreciate the discussions with Shane Smiskol at Openpilot's community.

\appendix

\section{Proof of the SS for the factory linear ACC planner} \label{factory-linear-SS-proof}
Here we show the linear ACC planner in \eqref{ACC-vt} is string stable for sensible values of $k$ and $\tau$.

\begin{proof} We first convert the time-domain planner trajectory in \eqref{ACC-vt} into frequency domain:
\begin{equation}
    v_{\text{target}}(j\omega) = k [x_{\text{lead}}(j\omega) - x_{\text{ego}}(j\omega) - \tau v_{\text{lead}}(j\omega)] + v_{\text{lead}}(j\omega)
    \label{linear_freq}
\end{equation}
where $\omega$ is the angular frequency, $j$ is the complex number indicator. As here we focus merely on the planner, we assume $v_{\text{ego}}=v_{\text{target}}$. Then substituting in \eqref{desire_d}, and applying the position as the integrator of corresponding speed: $x_{\text{lead}}(j\omega) = v_{\text{lead}}(j\omega)/j\omega$, and $x_{\text{ego}}(j\omega) = v_{\text{ego}}(j\omega)/j\omega = v_{\text{target}}(j\omega)/j\omega$, we can rearrange \eqref{linear_freq} into the well-known speed to speed transfer function used in SS analysis \citep{wilson2011car,montanino2021string}:
\begin{equation}
    \frac{v_{\text{target}}(j\omega)}{v_{\text{lead}}(j\omega)} = \frac{j(1-k \tau)\omega + k}{j\omega + k}
    \label{linearFreq}
\end{equation}
As $0 \le k \tau \le 2$ indicates $|1-k \tau| \le 1$, we can ensure that:
\begin{equation}
     \left \|\frac{v_{\text{target}}(j\omega)}{v_{\text{lead}}(j\omega)} \right\|^2 _2= \frac{[(1-k \tau)\omega]^2 + k ^2}{\omega^2 + k ^2} \le 1
    \label{linearSScond}
\end{equation}
which satisfies the frequency-domain SS criterion (\cite{naus2010string,Feng2019StringSF}): $\left \|\frac{v_{\text{target}}(j\omega)}{v_{\text{lead}}(j\omega)} \right\|_\infty \le 1$, indicating the string-stable property of the linear planner. Note that $\| v(j\omega) \|_\infty = \max _{\omega \in \mathbb{R}} v(j\omega)$, and $\| v(j\omega) \|_2 = \sqrt{\frac{1}{2\pi} \int_{-\infty}^{\infty} \text{Trace} [v(j\omega)^{\dagger}v(j\omega)] d\omega}$, $\dagger$ means the conjugate transpose.
\end{proof}

The above proof indicates the SS condition for the factory linear ACC planner is $0 \le k \tau \le 2$. Notice  that meaningfulvalues of $k$ are near 0.1 \citep{shi2021empirical} for different speed levels, and the largest headway observed in the literature \citep{li2021ACC} is around 1.7s. These values enable the SS condition to be satisfied naturally.

\bibliography{Reference}

\begin{thebibliography}{47}
\expandafter\ifx\csname natexlab\endcsname\relax\def\natexlab#1{#1}\fi
\providecommand{\url}[1]{\texttt{#1}}
\providecommand{\href}[2]{#2}
\providecommand{\path}[1]{#1}
\providecommand{\DOIprefix}{doi:}
\providecommand{\ArXivprefix}{arXiv:}
\providecommand{\URLprefix}{URL: }
\providecommand{\Pubmedprefix}{pmid:}
\providecommand{\doi}[1]{\href{http://dx.doi.org/#1}{\path{#1}}}
\providecommand{\Pubmed}[1]{\href{pmid:#1}{\path{#1}}}
\providecommand{\bibinfo}[2]{#2}
\ifx\xfnm\relax \def\xfnm[#1]{\unskip,\space#1}\fi
\bibitem[{Ahn et~al.(2004)Ahn, Cassidy and Laval}]{ahn2004verification}
\bibinfo{author}{Ahn, S.}, \bibinfo{author}{Cassidy, M.J.},
  \bibinfo{author}{Laval, J.}, \bibinfo{year}{2004}.
\newblock \bibinfo{title}{Verification of a simplified car-following theory}.
\newblock \bibinfo{journal}{Transportation Research Part B: Methodological}
  \bibinfo{volume}{38}, \bibinfo{pages}{431--440}.
\bibitem[{COLLINS(2021)}]{tunefast}
\bibinfo{author}{COLLINS, D.}, \bibinfo{year}{2021}.
\newblock \bibinfo{title}{How to tune servo systems for high dynamic response?}
\newblock \URLprefix
  \url{https://www.motioncontroltips.com/faq-tune-servo-system-high-dynamic-response/}.
\bibitem[{Comma.ai(2020)}]{comma2}
\bibinfo{author}{Comma.ai}, \bibinfo{year}{2020}.
\newblock \bibinfo{title}{Comma.ai two setup}.
\newblock \URLprefix \url{https://comma.ai/setup}.
\bibitem[{Comma.ai(2021)}]{Openpilot}
\bibinfo{author}{Comma.ai}, \bibinfo{year}{2021}.
\newblock \bibinfo{title}{Comma.ai – introducing openpilot}.
\newblock \URLprefix \url{https://comma.ai/}.
\bibitem[{Corona and De~Schutter(2008)}]{corona2008adaptive}
\bibinfo{author}{Corona, D.}, \bibinfo{author}{De~Schutter, B.},
  \bibinfo{year}{2008}.
\newblock \bibinfo{title}{Adaptive cruise control for a smart car: A comparison
  benchmark for mpc-pwa control methods}.
\newblock \bibinfo{journal}{IEEE Transactions on Control Systems Technology}
  \bibinfo{volume}{16}, \bibinfo{pages}{365--372}.
\bibitem[{Diehl()}]{Acado}
\bibinfo{author}{Diehl, M.}, .
\newblock \bibinfo{title}{Toolkit for automatic control and dynamic
  optimization}.
\newblock \URLprefix \url{https://acado.github.io/}.
\bibitem[{Eilbert et~al.(2020)Eilbert, Chouinard, Tiernan and
  Smith}]{Eilbert2020PerformanceCO}
\bibinfo{author}{Eilbert, A.}, \bibinfo{author}{Chouinard, A.M.},
  \bibinfo{author}{Tiernan, T.}, \bibinfo{author}{Smith, S.},
  \bibinfo{year}{2020}.
\newblock \bibinfo{title}{Performance comparisons of cooperative and adaptive
  cruise control testing}.
\bibitem[{Feng et~al.(2019)Feng, Zhang, Li, Cao, Liu and Li}]{Feng2019StringSF}
\bibinfo{author}{Feng, S.}, \bibinfo{author}{Zhang, Y.}, \bibinfo{author}{Li,
  S.}, \bibinfo{author}{Cao, Z.}, \bibinfo{author}{Liu, H.},
  \bibinfo{author}{Li, L.}, \bibinfo{year}{2019}.
\newblock \bibinfo{title}{String stability for vehicular platoon control:
  Definitions and analysis methods}.
\newblock \bibinfo{journal}{Annu. Rev. Control.} \bibinfo{volume}{47},
  \bibinfo{pages}{81--97}.
\bibitem[{Gong et~al.(2016)Gong, Shen and Du}]{Gong2016ConstrainedOA}
\bibinfo{author}{Gong, S.}, \bibinfo{author}{Shen, J.}, \bibinfo{author}{Du,
  L.}, \bibinfo{year}{2016}.
\newblock \bibinfo{title}{Constrained optimization and distributed computation
  based car following control of a connected and autonomous vehicle platoon}.
\newblock \bibinfo{journal}{Transportation Research Part B-methodological}
  \bibinfo{volume}{94}, \bibinfo{pages}{314--334}.
\bibitem[{Gunter et~al.(2020)Gunter, Gloudemans, Stern, McQuade, Bhadani,
  Bunting, Delle~Monache, Lysecky, Seibold, Sprinkle
  et~al.}]{gunter2020commercially}
\bibinfo{author}{Gunter, G.}, \bibinfo{author}{Gloudemans, D.},
  \bibinfo{author}{Stern, R.E.}, \bibinfo{author}{McQuade, S.},
  \bibinfo{author}{Bhadani, R.}, \bibinfo{author}{Bunting, M.},
  \bibinfo{author}{Delle~Monache, M.L.}, \bibinfo{author}{Lysecky, R.},
  \bibinfo{author}{Seibold, B.}, \bibinfo{author}{Sprinkle, J.}, et~al.,
  \bibinfo{year}{2020}.
\newblock \bibinfo{title}{Are commercially implemented adaptive cruise control
  systems string stable?}
\newblock \bibinfo{journal}{IEEE Transactions on Intelligent Transportation
  Systems} .
\bibitem[{Gunter et~al.(2019)Gunter, Janssen, Barbour, Stern and
  Work}]{gunter2019model}
\bibinfo{author}{Gunter, G.}, \bibinfo{author}{Janssen, C.},
  \bibinfo{author}{Barbour, W.}, \bibinfo{author}{Stern, R.E.},
  \bibinfo{author}{Work, D.B.}, \bibinfo{year}{2019}.
\newblock \bibinfo{title}{Model-based string stability of adaptive cruise
  control systems using field data}.
\newblock \bibinfo{journal}{IEEE Transactions on Intelligent Vehicles}
  \bibinfo{volume}{5}, \bibinfo{pages}{90--99}.
\bibitem[{Kanojiya and Meshram(2012)}]{kanojiya2012optimal}
\bibinfo{author}{Kanojiya, R.G.}, \bibinfo{author}{Meshram, P.},
  \bibinfo{year}{2012}.
\newblock \bibinfo{title}{Optimal tuning of pi controller for speed control of
  dc motor drive using particle swarm optimization}, in:
  \bibinfo{booktitle}{2012 international conference on advances in power
  conversion and energy technologies (APCET)}, \bibinfo{organization}{IEEE}.
  pp. \bibinfo{pages}{1--6}.
\bibitem[{Laval et~al.(2014)Laval, Toth and Zhou}]{laval2014parsimonious}
\bibinfo{author}{Laval, J.A.}, \bibinfo{author}{Toth, C.S.},
  \bibinfo{author}{Zhou, Y.}, \bibinfo{year}{2014}.
\newblock \bibinfo{title}{A parsimonious model for the formation of
  oscillations in car-following models}.
\newblock \bibinfo{journal}{Transportation Research Part B: Methodological}
  \bibinfo{volume}{70}, \bibinfo{pages}{228--238}.
\bibitem[{Li et~al.(2010)Li, Li, Rajamani and Wang}]{li2010model}
\bibinfo{author}{Li, S.}, \bibinfo{author}{Li, K.}, \bibinfo{author}{Rajamani,
  R.}, \bibinfo{author}{Wang, J.}, \bibinfo{year}{2010}.
\newblock \bibinfo{title}{Model predictive multi-objective vehicular adaptive
  cruise control}.
\newblock \bibinfo{journal}{IEEE Transactions on control systems technology}
  \bibinfo{volume}{19}, \bibinfo{pages}{556--566}.
\bibitem[{Li et~al.(2021a)Li, Chen, Zhou, Laval and Xie}]{li2021ACC}
\bibinfo{author}{Li, T.}, \bibinfo{author}{Chen, D.}, \bibinfo{author}{Zhou,
  H.}, \bibinfo{author}{Laval, J.}, \bibinfo{author}{Xie, Y.},
  \bibinfo{year}{2021}a.
\newblock \bibinfo{title}{Car-following behavior characteristics of adaptive
  cruise control vehicles based on empirical experiments}.
\newblock \bibinfo{journal}{Transportation Research Part B: Methodological}
  \bibinfo{volume}{147}, \bibinfo{pages}{67--91}.
\newblock \DOIprefix\doi{https://doi.org/10.1016/j.trb.2021.03.003}.
\bibitem[{Li et~al.(2021b)Li, Chen, Zhou, Laval and Xie}]{li2021FD}
\bibinfo{author}{Li, T.}, \bibinfo{author}{Chen, D.}, \bibinfo{author}{Zhou,
  H.}, \bibinfo{author}{Laval, J.}, \bibinfo{author}{Xie, Y.},
  \bibinfo{year}{2021}b.
\newblock \bibinfo{title}{On the fundamental diagrams of commercial adaptive
  cruise control: Worldwide experimental evidence}.
\newblock \bibinfo{journal}{arXiv preprint arXiv:2105.05380} .
\bibitem[{Li(2020)}]{li2020trade}
\bibinfo{author}{Li, X.}, \bibinfo{year}{2020}.
\newblock \bibinfo{title}{Trade-off between safety, mobility and stability in
  automated vehicle following control: An analytical method} .
\bibitem[{Liang and Peng(2000)}]{Liang2000StringSA}
\bibinfo{author}{Liang, C.}, \bibinfo{author}{Peng, H.}, \bibinfo{year}{2000}.
\newblock \bibinfo{title}{String stability analysis of adaptive cruise
  controlled vehicles}.
\newblock \bibinfo{journal}{Jsme International Journal Series C-mechanical
  Systems Machine Elements and Manufacturing} \bibinfo{volume}{43},
  \bibinfo{pages}{671--677}.
\bibitem[{Liang and Peng(1999)}]{liang1999optimal}
\bibinfo{author}{Liang, C.Y.}, \bibinfo{author}{Peng, H.},
  \bibinfo{year}{1999}.
\newblock \bibinfo{title}{Optimal adaptive cruise control with guaranteed
  string stability}.
\newblock \bibinfo{journal}{Vehicle system dynamics} \bibinfo{volume}{32},
  \bibinfo{pages}{313--330}.
\bibitem[{Lu and Shladover(2018)}]{Lu2018TruckCS}
\bibinfo{author}{Lu, X.}, \bibinfo{author}{Shladover, S.},
  \bibinfo{year}{2018}.
\newblock \bibinfo{title}{Truck cacc system designand dsrc messages}.
\bibitem[{Lykiardopulou(2021)}]{tesla-remove}
\bibinfo{author}{Lykiardopulou, I.}, \bibinfo{year}{2021}.
\newblock \bibinfo{title}{Tesla is removing radar from autopilot, and it makes
  absolutely no sense} \URLprefix
  \url{https://thenextweb.com/news/tesla-is-removing-radar-from-autopilot-and-it-makes-absolutely-no-sense}.
\bibitem[{Makridis et~al.(2021)Makridis, Mattas, Anesiadou and
  Ciuffo}]{makridis2021openacc}
\bibinfo{author}{Makridis, M.}, \bibinfo{author}{Mattas, K.},
  \bibinfo{author}{Anesiadou, A.}, \bibinfo{author}{Ciuffo, B.},
  \bibinfo{year}{2021}.
\newblock \bibinfo{title}{Openacc. an open database of car-following
  experiments to study the properties of commercial acc systems}.
\newblock \bibinfo{journal}{Transportation research part C: emerging
  technologies} \bibinfo{volume}{125}, \bibinfo{pages}{103047}.
\bibitem[{Montanino and Punzo(2021)}]{montanino2021string}
\bibinfo{author}{Montanino, M.}, \bibinfo{author}{Punzo, V.},
  \bibinfo{year}{2021}.
\newblock \bibinfo{title}{On string stability of a mixed and heterogeneous
  traffic flow: A unifying modelling framework}.
\newblock \bibinfo{journal}{Transportation Research Part B: Methodological}
  \bibinfo{volume}{144}, \bibinfo{pages}{133--154}.
\bibitem[{Naus et~al.(2008)Naus, Ploeg, Van De~Molengraft and
  Steinbuch}]{naus2008explicit}
\bibinfo{author}{Naus, G.}, \bibinfo{author}{Ploeg, J.}, \bibinfo{author}{Van
  De~Molengraft, R.}, \bibinfo{author}{Steinbuch, M.}, \bibinfo{year}{2008}.
\newblock \bibinfo{title}{Explicit mpc design and performance-based tuning of
  an adaptive cruise control stop-\&-go}, in: \bibinfo{booktitle}{2008 IEEE
  Intelligent Vehicles Symposium}, \bibinfo{organization}{IEEE}. pp.
  \bibinfo{pages}{434--439}.
\bibitem[{Naus et~al.(2010)Naus, Vugts, Ploeg, van De~Molengraft and
  Steinbuch}]{naus2010string}
\bibinfo{author}{Naus, G.J.}, \bibinfo{author}{Vugts, R.P.},
  \bibinfo{author}{Ploeg, J.}, \bibinfo{author}{van De~Molengraft, M.J.},
  \bibinfo{author}{Steinbuch, M.}, \bibinfo{year}{2010}.
\newblock \bibinfo{title}{String-stable cacc design and experimental
  validation: A frequency-domain approach}.
\newblock \bibinfo{journal}{IEEE Transactions on vehicular technology}
  \bibinfo{volume}{59}, \bibinfo{pages}{4268--4279}.
\bibitem[{Newell(2002)}]{newell2002simplified}
\bibinfo{author}{Newell, G.F.}, \bibinfo{year}{2002}.
\newblock \bibinfo{title}{A simplified car-following theory: a lower order
  model}.
\newblock \bibinfo{journal}{Transportation Research Part B: Methodological}
  \bibinfo{volume}{36}, \bibinfo{pages}{195--205}.
\bibitem[{O'Dwyer(2006)}]{o2006pi}
\bibinfo{author}{O'Dwyer, A.}, \bibinfo{year}{2006}.
\newblock \bibinfo{title}{Pi and pid controller tuning rules: an overview and
  personal perspective} .
\bibitem[{O'dwyer(2009)}]{handbook}
\bibinfo{author}{O'dwyer, A.}, \bibinfo{year}{2009}.
\newblock \bibinfo{title}{Handbook of PI and PID controller tuning rules}.
\newblock \bibinfo{publisher}{World Scientific}.
\bibitem[{Pancotti(2021)}]{honda-bosch}
\bibinfo{author}{Pancotti, J.}, \bibinfo{year}{2021}.
\newblock \bibinfo{title}{Honda bosch interface for openpilot}.
\newblock \URLprefix
  \url{https://github.com/jpancotti/openpilot/commit/468b89a}.
\bibitem[{Ploeg et~al.(2011)Ploeg, Scheepers, van Nunen, van~de Wouw and
  Nijmeijer}]{6082981}
\bibinfo{author}{Ploeg, J.}, \bibinfo{author}{Scheepers, B.T.M.},
  \bibinfo{author}{van Nunen, E.}, \bibinfo{author}{van~de Wouw, N.},
  \bibinfo{author}{Nijmeijer, H.}, \bibinfo{year}{2011}.
\newblock \bibinfo{title}{Design and experimental evaluation of cooperative
  adaptive cruise control}, in: \bibinfo{booktitle}{2011 14th International
  IEEE Conference on Intelligent Transportation Systems (ITSC)}, pp.
  \bibinfo{pages}{260--265}.
\newblock \DOIprefix\doi{10.1109/ITSC.2011.6082981}.
\bibitem[{Rakha et~al.(2001)Rakha, Lucic, Demarchi, Setti and
  Aerde}]{rakha2001vehicle}
\bibinfo{author}{Rakha, H.}, \bibinfo{author}{Lucic, I.},
  \bibinfo{author}{Demarchi, S.H.}, \bibinfo{author}{Setti, J.R.},
  \bibinfo{author}{Aerde, M.V.}, \bibinfo{year}{2001}.
\newblock \bibinfo{title}{Vehicle dynamics model for predicting maximum truck
  acceleration levels}.
\newblock \bibinfo{journal}{Journal of transportation engineering}
  \bibinfo{volume}{127}, \bibinfo{pages}{418--425}.
\bibitem[{{Shaw} and {Hedrick}(2007)}]{4282789}
\bibinfo{author}{{Shaw}, E.}, \bibinfo{author}{{Hedrick}, J.K.},
  \bibinfo{year}{2007}.
\newblock \bibinfo{title}{String stability analysis for heterogeneous vehicle
  strings}, in: \bibinfo{booktitle}{2007 American Control Conference}, pp.
  \bibinfo{pages}{3118--3125}.
\newblock \DOIprefix\doi{10.1109/ACC.2007.4282789}.
\bibitem[{Shi and Li(2021)}]{shi2021empirical}
\bibinfo{author}{Shi, X.}, \bibinfo{author}{Li, X.}, \bibinfo{year}{2021}.
\newblock \bibinfo{title}{Empirical study on car-following characteristics of
  commercial automated vehicles with different headway settings}.
\newblock \bibinfo{journal}{Transportation Research Part C: Emerging
  Technologies} \bibinfo{volume}{128}, \bibinfo{pages}{103134}.
\bibitem[{Shladover(2009)}]{Shladover2009EffectsOC}
\bibinfo{author}{Shladover, S.}, \bibinfo{year}{2009}.
\newblock \bibinfo{title}{Effects of cooperative adaptive cruise control on
  traffic flow: Testing drivers' choices of following distances}.
\newblock \bibinfo{journal}{PATH research report} .
\bibitem[{Smiskol(2021)}]{compute_gb}
\bibinfo{author}{Smiskol, S.}, \bibinfo{year}{2021}.
\newblock \bibinfo{title}{Discussion on the "compute\_gb" function in openpilot
  community}.
\bibitem[{Wang and Shao(2000)}]{wang2000optimal}
\bibinfo{author}{Wang, Y.G.}, \bibinfo{author}{Shao, H.H.},
  \bibinfo{year}{2000}.
\newblock \bibinfo{title}{Optimal tuning for pi controller}.
\newblock \bibinfo{journal}{Automatica} \bibinfo{volume}{36},
  \bibinfo{pages}{147--152}.
\bibitem[{Willis(1999)}]{willis1999proportional}
\bibinfo{author}{Willis, M.}, \bibinfo{year}{1999}.
\newblock \bibinfo{title}{Proportional-integral-derivative control}.
\newblock \bibinfo{journal}{Dept. of Chemical and Process Engineering
  University of Newcastle} .
\bibitem[{Wilson and Ward(2011)}]{wilson2011car}
\bibinfo{author}{Wilson, R.E.}, \bibinfo{author}{Ward, J.A.},
  \bibinfo{year}{2011}.
\newblock \bibinfo{title}{Car-following modelsfifty years of linear stability
  analysis a mathematical perspective}.
\newblock \bibinfo{journal}{Transportation Planning and Technology}
  \bibinfo{volume}{34}, \bibinfo{pages}{3--18}.
\bibitem[{Xu and Laval(2020)}]{xu2020statistical}
\bibinfo{author}{Xu, T.}, \bibinfo{author}{Laval, J.}, \bibinfo{year}{2020}.
\newblock \bibinfo{title}{Statistical inference for two-regime stochastic
  car-following models}.
\newblock \bibinfo{journal}{Transportation Research Part B: Methodological}
  \bibinfo{volume}{134}, \bibinfo{pages}{210--228}.
\bibitem[{Yanakiev and Kanellakopoulos(1995)}]{yanakiev}
\bibinfo{author}{Yanakiev, D.}, \bibinfo{author}{Kanellakopoulos, I.},
  \bibinfo{year}{1995}.
\newblock \bibinfo{title}{Variable time headway for string stability of
  automated heavy-duty vehicles}, in: \bibinfo{booktitle}{Proceedings of 1995
  34th IEEE Conference on Decision and Control}, \bibinfo{organization}{IEEE}.
  pp. \bibinfo{pages}{4077--4081}.
\bibitem[{Zheng et~al.(2017)Zheng, Li, Li, Borrelli and Hedrick}]{7546918}
\bibinfo{author}{Zheng, Y.}, \bibinfo{author}{Li, S.E.}, \bibinfo{author}{Li,
  K.}, \bibinfo{author}{Borrelli, F.}, \bibinfo{author}{Hedrick, J.K.},
  \bibinfo{year}{2017}.
\newblock \bibinfo{title}{Distributed model predictive control for
  heterogeneous vehicle platoons under unidirectional topologies}.
\newblock \bibinfo{journal}{IEEE Transactions on Control Systems Technology}
  \bibinfo{volume}{25}, \bibinfo{pages}{899--910}.
\newblock \DOIprefix\doi{10.1109/TCST.2016.2594588}.
\bibitem[{Zhou et~al.(2020)Zhou, Gong, Wang and
  Peeta}]{Zhou2020SmoothSwitchingCC}
\bibinfo{author}{Zhou, A.}, \bibinfo{author}{Gong, S.}, \bibinfo{author}{Wang,
  C.}, \bibinfo{author}{Peeta, S.}, \bibinfo{year}{2020}.
\newblock \bibinfo{title}{Smooth-switching control-based cooperative adaptive
  cruise control by considering dynamic information flow topology}.
\newblock \bibinfo{journal}{Transportation Research Record}
  \bibinfo{volume}{2674}, \bibinfo{pages}{444 -- 458}.
\bibitem[{Zhou(2020)}]{zhou-mpc3}
\bibinfo{author}{Zhou, H.}, \bibinfo{year}{2020}.
\newblock \bibinfo{title}{The mpc designed for end-to-end longitudinal
  self-driving at openpilot, comma.ai.}
\newblock \URLprefix \url{https://howardchow92.medium.com}.
\bibitem[{Zhou et~al.(2021)Zhou, Laval, Zhou, Wang, Wu, Qing and
  Peeta}]{zhou2021review}
\bibinfo{author}{Zhou, H.}, \bibinfo{author}{Laval, J.A.},
  \bibinfo{author}{Zhou, A.}, \bibinfo{author}{Wang, Y.}, \bibinfo{author}{Wu,
  W.}, \bibinfo{author}{Qing, Z.}, \bibinfo{author}{Peeta, S.},
  \bibinfo{year}{2021}.
\newblock \bibinfo{title}{Review of learning-based longitudinal motion planning
  for autonomous vehicles: Implications on traffic congestion}.
\newblock \bibinfo{journal}{Transportation Research Board 100th Annual
  MeetingTransportation Research Board} .
\bibitem[{Zhou and Peng(2005)}]{zhou2005range}
\bibinfo{author}{Zhou, J.}, \bibinfo{author}{Peng, H.}, \bibinfo{year}{2005}.
\newblock \bibinfo{title}{Range policy of adaptive cruise control vehicles for
  improved flow stability and string stability}.
\newblock \bibinfo{journal}{IEEE Transactions on intelligent transportation
  systems} \bibinfo{volume}{6}, \bibinfo{pages}{229--237}.
\bibitem[{Zhou and Ahn(2019)}]{zhou2019robust}
\bibinfo{author}{Zhou, Y.}, \bibinfo{author}{Ahn, S.}, \bibinfo{year}{2019}.
\newblock \bibinfo{title}{Robust local and string stability for a decentralized
  car following control strategy for connected automated vehicles}.
\newblock \bibinfo{journal}{Transportation Research Part B: Methodological}
  \bibinfo{volume}{125}, \bibinfo{pages}{175--196}.
\bibitem[{Ziegler and Nichols(1942)}]{Ziegler1942OptimumSF}
\bibinfo{author}{Ziegler, J.G.}, \bibinfo{author}{Nichols, N.},
  \bibinfo{year}{1942}.
\newblock \bibinfo{title}{Optimum settings for automatic controllers}.
\newblock \bibinfo{journal}{Journal of Dynamic Systems Measurement and
  Control-transactions of The Asme} \bibinfo{volume}{115},
  \bibinfo{pages}{220--222}.

\end{thebibliography}

\end{document}